\documentclass[11pt,a4paper]{article}
\pdfoutput=1
\usepackage{jcappub}
\usepackage[T1]{fontenc}
\title{Axions in Neutron Star Mergers}

\author[a]{Steven P. Harris,}
\author[b]{Jean-Fran\c cois Fortin,}
\author[c]{Kuver Sinha,}
\author[a]{Mark G. Alford}

\affiliation[a]{Physics Department, Washington University, St.~Louis, MO~63130, USA}
\affiliation[b]{D\'epartement de Physique, de G\'enie Physique et d'Optique, Universit\'e Laval, Qu\'ebec, QC G1V 0A6, Canada}
\affiliation[c]{Department of Physics and Astronomy, University of Oklahoma, Norman, OK 73019, USA}

\emailAdd{harrissp@wustl.edu}
\emailAdd{jean-francois.fortin@phy.ulaval.ca}
\emailAdd{kuver.sinha@ou.edu}
\emailAdd{alford@physics.wustl.edu}

\abstract{Supernovae and cooling neutron stars have long been used to constrain the properties of axions, such as their mass and interactions with nucleons and other Standard Model particles. We investigate the prospects of using neutron star mergers as a similar location where axions can be probed in the future. We examine the impact axions would have on mergers, considering both the possibility that they free-stream through the dense nuclear matter and the case where they are trapped.  We calculate the mean free path of axions in merger conditions, and find that they would free-stream through the merger in all thermodynamic conditions. In contrast to previous calculations, we integrate over the entire phase space while using a relativistic treatment of the nucleons, assuming the matrix element is momentum-independent.  In particular, we use a relativistic mean field theory to describe the nucleons, taking into account the precipitous decrease in the effective mass of the nucleons as density increases above nuclear saturation density. We find that within current constraints on the axion-neutron coupling, axions could cool nuclear matter on timescales relevant to neutron star mergers. Our results may be regarded as first steps aimed at understanding how axions  affect merger simulations and potentially interface with observations.}
\begin{document}
\maketitle
\flushbottom
%%%%%%%%%%%%%%%%%%%%%%%%%%%%%%%%%%%%%%%
\section{Introduction}
\label{sec:intro}
The growing importance of hidden sectors populated by low-mass particles has led to a renewed interest in astrophysical probes of such species. Traditionally supernovae have served as important laboratories where light species (such as axions and axion-like particles, ALPs) can be gainfully constrained (we refer to \cite{Chang:2018rso} for a recent status of this frontier, and \cite{Raffelt:1996wa} for a classic review). 

Our purpose in this paper is to ask the following question: can the growing data for neutron star mergers be leveraged as a similar location where ALPs, and perhaps light hidden species more generally, can be probed in the future? More concretely, we take the first steps to initiate a program aimed at understanding how ALPs may affect merger simulations and interface with observations.

The detection of gravitational waves from the  binary neutron star merger event GW170817 by the LIGO/Virgo collaboration, and the observation of correlated electromagnetic signals,  has ushered in the era of multi-messenger astronomy \cite{TheLIGOScientific:2017qsa}. 

Simulations using numerical relativity of equal and unequal mass binaries for several equations of state have shown that if the total mass of the system is not too large, the merger could produce a remnant neutron star that lives anywhere from tens of milliseconds to tens of seconds, perhaps more \cite{Lucca:2019ohp,Baiotti:2016qnr}. In addition, Gill \textit{et.~al} used properties of the kilonova and the delay time between the arrival of the gravitational wave signal and the gamma ray burst to conjecture that GW170817 lead to a hypermassive neutron star that survived for approximately one second \cite{Gill:2019bvq}.  Several studies have focused on the internal structure of the hypermassive neutron star and its temperature and density distributions. We refer to \cite{Hanauske:2019qgs} for a recent review of these topics.

The simulation of merging compact objects is a highly complex subject that incorporates nuclear reactions, magnetohydrodynamics, a hot nuclear equation of state (EoS), and the effects of general relativity \cite{Radice:2018pdn, Radice:2016rys, Wanajo:2014wha, Kasen:2013xka, Metzger:2017wot}. In addition, a critical role is played by neutrino transport, which determines properties of the ejected material such as their brightness and color. The internal region of the hypermassive neutron star can reach densities of several times nuclear-matter saturation density and temperatures of order tens of MeV. Increasingly sophisticated simulations are investigating the evolution of this system and its post-merger history, which depends on the mass, EoS, and the strength of the magnetic fields \cite{Shibata:2006nm, Baiotti:2008ra, Hotokezaka:2011dh, Baiotti:2016qnr}.

The addition of new physics in the form of axions or ALPs will have important consequences on all aspects of the physics of neutron star mergers. While incorporating the full effect of ALPs in merger simulations is clearly a difficult problem, a preliminary effort has been made by Dietrich and Clough \cite{Dietrich:2019shr}.  They model axion cooling of a merger by using standard axion emissivity expressions from Brinkmann and Turner \cite{Brinkmann:1988vi} in nuclear matter in both the non-degenerate and degenerate regimes.

To determine the role of axions in neutron star mergers, we consider a couple of possibilities.  If axions have a long mean free path (MFP) compared to the size of the merger (20-30 kilometers in diameter \cite{Baiotti:2016qnr,Radice:2016dwd}), then they free-stream through the nuclear matter, taking energy away from the merger which results in cooling.  On the other hand, if axions have a relatively short mean free path they would contribute to transport inside the merger.  

To calculate the mean free path we first discuss the production (or absorption) of ALPs (with field operator $a$) via bremsstrahlung from neutrons
(with field operator $\psi_n$). The relevant coupling term in the
Lagrangian is $\mathcal{L} = G_{an}(\partial_{\mu}a)\,\bar{\psi}_n \gamma^{\mu}\gamma_5\psi_n$.  The standard calculations of axion mean free paths and emissivities rely on the Fermi surface (FS) approximation: we propose an improvement to the Fermi surface approximation which extends its validity to semi-degenerate nuclear matter.  Our main result is a calculation of the axion emissivity and mean free path, where the only approximation is the
assumption of a momentum-independent matrix element for the neutron bremsstrahlung process.  We keep the relativistic energy dispersion of the neutrons and we keep the axion momentum in the energy-momentum conserving delta function.  The full phase space integration is valid for degenerate neutrons as well as non-degenerate neutrons.  We then discuss the ALP-transparent regime (where the axion mean free path is comparable to or larger than the system size) and ALP-trapped regime of the merger (where the axion mean free path is much less than the system size). For the former, the temperature of fluid elements radiating ALPs as a function of time is computed and the characteristic cooling times are obtained in Figs.~\ref{fig:radiative_cooling_dens_temp} and \ref{fig:rad_cool_time_vs_G}. In the trapped regime, the timescale of thermal equilibration for a fluid element to transfer heat to its neighboring fluid elements is computed. The results are depicted in Fig.~\ref{fig:conductivecooling}. 

Throughout our work, we show the constraints coming from SN1987A on the axion-neutron coupling constant $G_{an}$ \cite{Graham:2015ouw,Raffelt:1996wa,PhysRevD.42.3297,PhysRevD.39.1020,Mayle:1987as,Mayle:1989yx,PhysRevLett.60.1793,Turner:1987by} and discuss the interplay of our results with those coming from supernova physics.  The limit obtained by SN1987A alone cannot rule out a very strong coupling between axions and neutrons, such that axions would be trapped within the supernova.  However, strong couplings are ruled out by cosmological arguments (see \cite{Tanabashi:2018oca} for a review).  Since supernova bounds can vary considerably depending on the details of the core-collapse simulations (we refer to \cite{Bar:2019ifz} for a recent critical assessment), our approach is to treat the SN1987A constraint ($G_{an}\lesssim 7.9\times 10^{-10}\text{ GeV}^{-1}$) loosely, and thus we examine a range of axion-neutron couplings that extends somewhat above the upper bound, down to significantly below the upper bound.

We note that there has been substantial recent work in related directions, mainly focused on the mass regime of the QCD axion where its Compton wavelength is comparable to the binary star  separation. Then, long-range axionic forces may be operational between neutron stars \cite{Hook:2017psm, Huang:2018pbu, Poddar:2019zoe}. This is an important direction independent of the one we are pursuing, and holds the promise of probing axions using gravitational wave emission from neutron star mergers, and perhaps associated electromagnetic signals as well. 

The interplay of the post-merger history with the gravitational wave signal is an active area of study \cite{Gill:2019bvq} and is also relevant to the axion mass regimes we explore, albeit somewhat indirectly. Our approach is to consider the emission of ALPs by bremsstrahlung; indeed, if the viscous dissipation and energy transport properties of the merger product are relevant on time scales of order $10$\,ms, these properties can leave imprints on the gravitational wave signal. Conversely, the gravitational wave signals can be used to illuminate these properties \cite{Alford:2017rxf}. Thus, the careful treatment of ALPs in the merger environment which we initiate here may ultimately also lead to correlated gravitational wave and electromagnetic signals.

We work in natural units, where $\hbar=c=k_B=1$.  
%%%%%%%%%%%%%%%%%%%%%%%%%%%%%%%%%%%%%%%%%%%%%%%%%%
\section{Axion production in nuclear matter}
\label{sec:axions_in_nuclear_matter}
Axions are proposed to couple to neutrons with the interaction term $\mathcal{L} = G_{an}\partial_{\mu}a\bar{\psi}_n\gamma^{\mu}\gamma_5\psi_n$ \cite{Brinkmann:1988vi}.  A neutron by itself cannot emit an axion because of energy-momentum conservation, so a spectator nucleon is required to donate energy/momentum to the processes to allow it to proceed.  The strong interaction between the spectator nucleon and the nucleon emitting the axion (throughout this paper, we assume that both nucleons are neutrons) is modeled by one-pion exchange (OPE) \cite{OPE} with Lagrangian  $\mathcal{L}_{n\pi} = i (2m_n/m_{\pi}) f \gamma^5 \pi_0 \bar{\psi}_n\psi_n$, where $f\approx 1$.  The neutron and pion masses here are their respective masses in vacuum.  Thus, axions can be created and absorbed by the neutron bremsstrahlung process $n + n \leftrightarrow n + n + a$.  This process is described at tree level by eight Feynman diagrams (see Fig.~4 of Ref.~\cite{Brinkmann:1988vi}), giving rise to the matrix element (derived in the appendix of \cite{Brinkmann:1988vi})
\begin{equation}
    S\sum_{\text{spins}} \vert \mathcal{M}\vert^2 = \frac{256}{3}\frac{f^4m_n^4G_{an}^2}{m_{\pi}^4}\left[ \frac{\mathbf{k}^4}{\left(\mathbf{k}^2+m_{\pi}^2\right)^2}+\frac{\mathbf{l}^4}{\left(\mathbf{l}^2+m_{\pi}^2\right)^2}+\frac{\mathbf{k}^2\mathbf{l}^2-3\left(\mathbf{k}\cdot\mathbf{l}\right)^2}{\left(\mathbf{k}^2+m_{\pi}^2\right)\left(\mathbf{l}^2+m_{\pi}^2\right)}    \right],\label{eq:matrix_element}
\end{equation}
where $\mathbf{k}$ and $\mathbf{l}$ are three-momentum transfers $\mathbf{k}=\mathbf{p_2}-\mathbf{p_4}$ and $\mathbf{l}=\mathbf{p_2}-\mathbf{p_3}$.  The symmetry factor for these diagrams is $S=1/4$, due to the presence of two identical particles in both the initial and final states.  As above, the prefactors of the neutron mass $m_n$ and pion mass $m_{\pi}$ correspond to respective masses in vacuum, since they arise from the definitions of the couplings in the pion-neutron Lagrangian shown above.  The dot product term in the matrix element is often written as $\beta \equiv 3\langle (\mathbf{\hat{k}}\cdot\mathbf{\hat{l}})^2\rangle$, where the brackets denote an average over phase space, which is a common technique to simplify the matrix element \cite{Brinkmann:1988vi}.

For a QCD axion, the axion mass $m_a$ is related to the axion-neutron coupling strength $G_{an}$ through\footnote{This expression comes from Eq.~(3.2) in \cite{Raffelt:2006cw}, where $f_a$ is found in terms of $G_{an}$ by matching the coefficients of the axion-neutron interaction term in the Lagrangian (given in Sec.~\ref{sec:intro} of our paper and Eq.~(3.6) in Ref.~\cite{Raffelt:2006cw}, taking $C_j \approx 1$.)  }
\begin{equation}
   m_a = 1.2\times 10^7 \text{ eV } \left(\frac{G_{an}}{\text{GeV}^{-1}}\right)
\end{equation}
Given current constraints on the axion-neutron coupling, the mass of the QCD axion must be well below 1 eV, which is much less than the typical momentum scales of order 100 MeV in neutron stars, thus we treat all ALPs as ultrarelativistic particles in our calculations.

We model the nuclear matter inside a neutron star with the $NL\rho$ EoS \cite{Liu:2001iz}, which is a relativistic mean field theory where nucleons interact by exchanging mesons, namely, the scalar $\sigma$ meson and the $\omega$ and $\rho$ vector mesons.  This EoS supports a $2M_{\odot}$ neutron star and has a pressure consistent with GW170817 and NICER data (see \cite{Essick:2020flb} for posterior distributions of the pressure at $0.5, 1, 2, 3$ times nuclear saturation density). In the mean field approximation, the neutron and proton behave like free particles with effective (Dirac) masses $m_* = m-g_{\sigma}\sigma$ and with effective chemical potentials $\mu_n^*=\mu_n-U_n$ and $\mu_p^*=\mu_p-U_p$, where $U_i$ are the nuclear mean fields
\begin{align}
U_n &= g_{\omega}\omega_0-\frac{1}{2}g_{\rho}\rho_{03}\label{eq:Un}\\
U_p &= g_{\omega}\omega_0+\frac{1}{2}g_{\rho}\rho_{03}\label{eq:Up}.
\end{align}
We use the relativistic definition of the chemical potentials $\mu_i$ and $\mu_i^*$, that is, they contain the rest mass of the particle.  The energy dispersion relations, modified by the presence of the nuclear mean field, become
\begin{align}
    E_n &= \sqrt{p^2+m_*^2}+U_n\label{eq:En}\\
    E_p &= \sqrt{p^2+m_*^2}+U_p.\label{eq:Ep}
\end{align}
Note that the $\rho$ meson distinguishes the neutron from the proton by creating a difference in mean field experienced by the neutron and proton.

The formalism for calculating the rate of a particle process in such a relativistic mean field theory is detailed in \cite{Fu:2008zzg}, which uses parameter set I of the model in \cite{Liu:2001iz}.  In the mean free path and emissivity calculations, $E^* \equiv \sqrt{p^2+m_*^2}$ should be used for the energies in the matrix element and in the energy factors in the denominator, while $E = E^* + U_n$ should be used in the energy delta function and the Fermi-Dirac factors \cite{Leinson:2002bw,Roberts:2016mwj,Fu:2008zzg}
\begin{equation}
    f_i = (1+e^{(E_i-\mu_n)/T})^{-1}.\label{eq:FD}
\end{equation}
Note that $E-\mu_n = E^*-\mu_n^*$.

\begin{figure}[tbp]
\centering
\includegraphics[width=.45\textwidth]{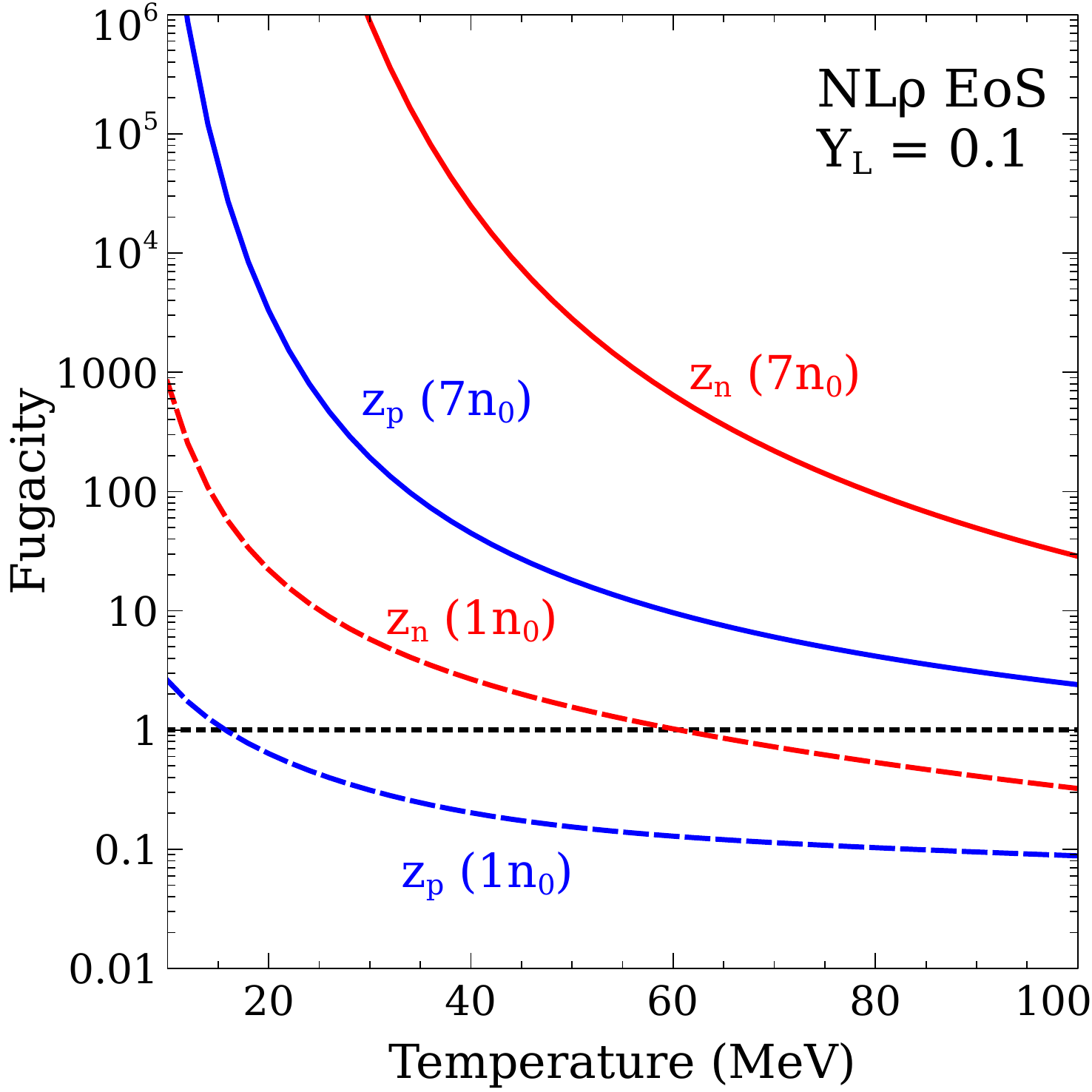}
\caption{Fugacities $z_i = e^{(\mu_i^*-m_{*,i})/T}$ of neutrons and protons in the $NL\rho$ EoS, with $Y_L = 0.1$.  A fugacity much larger than one indicates strongly degenerate particles, while a fugacity much smaller than one indicates non-degenerate particles.}
\label{fig:fugacity}
\end{figure}

In our calculations, we consider a lepton fraction of $Y_l = (n_{\nu}+n_e)/n_B = 0.1$, as this is a typical value for the neutrino-trapped region of a neutron star merger \cite{Alford:2019kdw} (though even this might be an overestimate \cite{Perego:2019adq,Most:2019onn}).  In Fig.~\ref{fig:fugacity}, we show the fugacities $z_i  = e^{(\mu_i^*-m_{*,i})/T}$ of neutrons and protons in this EoS at $Y_L = 0.1$.  A fugacity much larger than one indicates a strongly degenerate Fermi gas, while a fugacity much smaller than one indicates a highly non-degenerate Fermi gas \cite{Horowitz:2005zv,Fore:2019wib}.  Fig.~\ref{fig:fugacity} indicates that at
nuclear saturation density $n_0$, the protons are nondegenerate for nearly all considered temperatures, while the neutrons transition from degenerate to nondegenerate as the temperature goes above 50-60 MeV.  At $7n_0$, both types of nucleons are degenerate for all considered temperatures.

%%%%%%%%%%%%%%%%%%%%%%%%%%%%%%%%%%%%%%%%%%%%%%%%%%%%%%%%%
\section{Axion mean free path}
\label{sec:axion_mfp_intro}
The mean free path of an axion through nuclear matter depends on the temperature and density of the nuclear matter, but also the axion energy.  Often, we will consider axions with energy $\omega \approx 3T$, called ``thermal axions'', because this is the average energy of axions emitted via $n+n\rightarrow n+n+a$ from a fluid element of temperature $T$ \cite{Ishizuka:1989ts}.  In this section, we will compute the mean free path of axions, specializing to the case $\omega=3T$, and categorize thermodynamic conditions as either trapping axions or as allowing axions to free-stream.  As we are interested in neutron-star sized systems, we compare the mean free path to the system size, which is 20-30 kilometers in diameter.  If the mean free path of axions is less than 100 meters, we will consider them trapped, and if the mean free path is longer than 1 kilometer, then we will consider them free streaming.  These choices are somewhat arbitrary, and deserve further study.  Also, the intermediate region (mean free paths from 100\,m to 1\,km) is difficult to treat, as axions are neither trapped, forming a Bose sea, nor do they escape cleanly from the nuclear matter.

The mean free path $\lambda$ of an axion with energy $\omega$, due to absorption via $n+n+a\rightarrow n + n$, is given by \cite{PhysRevD.42.3297}
\begin{align}
&\lambda^{-1} = \int \frac{\mathop{d^3p}_1}{\left(2\pi\right)^3}\frac{\mathop{d^3p}_2}{\left(2\pi\right)^3}\frac{\mathop{d^3p}_3}{\left(2\pi\right)^3}\frac{\mathop{d^3p}_4}{\left(2\pi\right)^3}\frac{S\sum \vert \mathcal{M}\vert^2}{2^5 E_1^*E_2^*E_3^*E_4^*\omega}\label{eq:MFP_integral}\\
&\times \left(2\pi\right)^4\delta^4(p_1+p_2-p_3-p_4+\omega)  f_1f_2(1-f_3)(1-f_4).\nonumber
\end{align}
In the rest of this section, we will describe the results of calculating this MFP in various approximations, leaving the details to the appendix.

%%%%%%%%%%%%%%%%%%%%%%%%%%%%%%%%%%%%%%%%%
\subsection{Relativistic, arbitrary degeneracy}
\label{sec:axion_mfp_rel}
While at nuclear saturation density the nucleon effective mass is about 3/4 of its vacuum value \cite{glendenning2000compact}, at high baryon densities the nucleon effective mass decreases to only a few hundred MeV, which is comparable to or even lower than the typical momentum values of the nucleons participating in bremsstrahlung.  For example, the NL$\rho$ EoS predicts a neutron Dirac effective mass of 228\,MeV at a density of $7n_0$, where the neutron Fermi momentum is 604\,MeV.  Thus, it is important to use the full dispersion relation [Eq.~(\ref{eq:En})] for the neutrons in the MFP calculation at high densities.  We are able to do the phase space integration while using the relativistic neutron dispersion relation, provided we assume the matrix element is momentum-independent (a common, though not always necessary, approximation assumed in the literature by \cite{Brinkmann:1988vi,PhysRevD.42.3297,Paul:2018msp,Lee:2018lcj}).  The matrix element becomes independent of momentum if we assume that the momentum transfer magnitudes $\mathbf{k}^2$ and $\mathbf{l}^2$ have some typical value $k_{\text{typ}}^2$.  Then the matrix element can be written as 
\begin{equation}
    S\sum_{\text{spins}}\vert\mathcal{M}\vert^2 \approx 256\frac{f^4m_n^4G_{an}^2}{m_{\pi}^4}\left(1-\frac{\beta}{3}\right)\left(1+\frac{m_{\pi}^2}{k_{\text{typ}}^2}\right)^{-2}.\label{eq:cst_matrix}
\end{equation}
The typical values of momentum transfer are $k_{\text{typ}} \sim \sqrt{3m_*T}$ in the non-degenerate regime and $k_{\text{typ}}\sim p_{Fn}$ in the degenerate regime.  For temperatures and densities where the $NL\rho$ EoS predicts degenerate neutrons, $k_{\text{typ}}\sim p_{Fn}$ takes values between 320-600 MeV, while where the EoS predicts non-degenerate neutrons, $k_{\text{typ}}\sim\sqrt{3m_*T}$ takes values between 375-470 MeV.  Thus the factor of $\left(1+m_{\pi}^2/k_{\text{typ}}^2\right)^{-2}$ ranges from 0.78 to 0.91.  In the degenerate regime, $\beta = 0$, while in the non-degenerate regime, $\beta \approx 1.0845$ \cite{Brinkmann:1988vi}.

The momentum-independent matrix element can be pulled out of the phase space integral, and now the integral can be reduced to a 6-dimensional integral to be done numerically.  In addition to using the relativistic dispersion relation for neutrons, our calculation is also novel in that it keeps the axion three-momentum in the momentum-conserving delta function.  Finally, we emphasize that this approach to the mean free path integral is valid for arbitrary neutron degeneracy, as we make no simplifications to the Fermi-Dirac factors.  Our final expression for the axion mean free path in the constant-matrix-element approximation is given in Eq.~(\ref{eq:MFPexactanswer}), and the details of the calculation are given in Appendix \ref{sec:rel_PS_MFP}.

\begin{figure}[tbp]
\centering
\includegraphics[width=.45\textwidth]{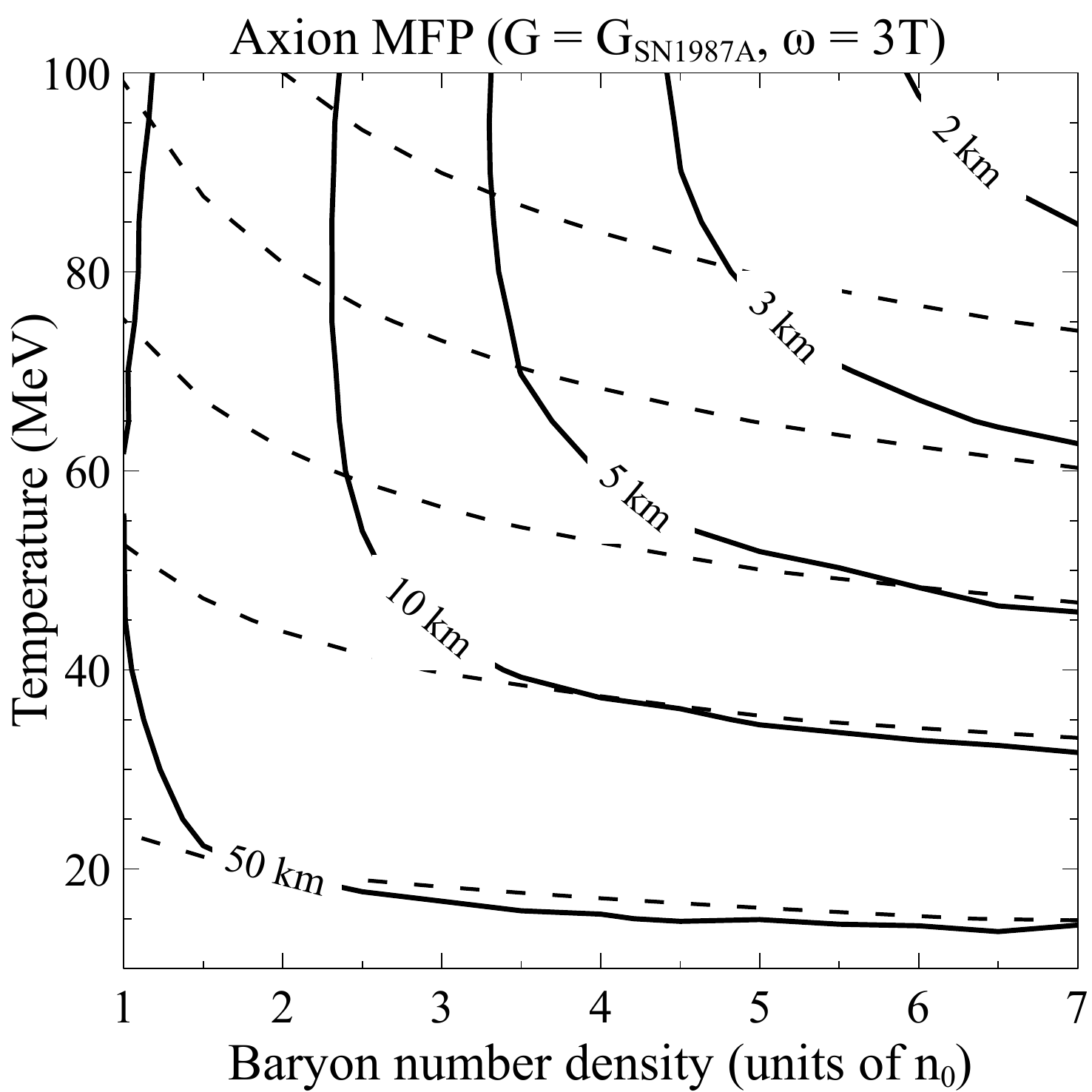}
\caption{Axion mean free path (for axions with energy $\omega = 3T$) due to absorption via $n+n+a\rightarrow n+n$ with an assumed axion-neutron coupling constant equal to the upper bound set by SN1987A.  The dashed contours correspond to the axion MFP calculated in the Fermi surface approximation, while the solid contours use the constant matrix element approximation of the axion mean free path.}
\label{fig:axionMFP}
\end{figure}

In Fig.~\ref{fig:axionMFP} we show a contour plot of the axion mean free path [Eq.~(\ref{eq:MFPexactanswer})].  We have assumed the axion energy $\omega = 3T$, and so Fig.~\ref{fig:axionMFP} does not depict the MFP of a fixed-energy axion, but of axions with progressively higher energies as the temperature increases.  The solid contours are the result of our constant-matrix-element approximation of the axion mean free path, where we have chosen $\beta=0$ for convenience.  The MFP is inversely proportional to the square of the unknown axion-neutron coupling constant, so we have chosen that coupling to be equal to the upper bound set by SN1987A, thus Fig.~\ref{fig:axionMFP} represents the smallest MFP allowed by SN1987A.  First, we see that the axion MFP is longer than 1 km for all thermodynamic conditions, indicating that axions will free-stream from the neutron star merger in which they are created.  Second, we see that the axion mean free path shrinks as matter becomes hotter and denser.  We see that at large neutron degeneracy (high density, low temperature), the mean free path of a thermal axion ($\omega=3T$) becomes relatively independent of density.

In Appendix \ref{sec:MFP_nonrel_PS}, we also present the phase space integral of the MFP while assuming non-relativistic neutrons.  In this case, the momentum dependence of the matrix element can be retained, and it is left inside the integral.  This calculation has been done before in the literature, but we present a version of the calculation well adapted to a relativistic mean field theory.
%%%%%%%%%%%%%%%%%%%%%%%%%%%%%%%%%%%%%%%%%%%%%%%%%
\subsection{Fermi surface approximation (degenerate neutrons)}
\label{sec:axion_mfp_FS}
The most common approximation of the full mean free path integral Eq.~(\ref{eq:MFP_integral}) is to assume the neutrons are strongly degenerate.  As can be seen in Fig.~\ref{fig:fugacity}, this assumption is valid at high densities like $7n_0$ for all temperatures encountered in neutron star mergers, but also even at lower densities like $n_0$, provided the temperature is below about 50 MeV.  We call this approximation the ``Fermi surface approximation'' because in degenerate nuclear matter only the particles near the Fermi surface can participate in any reactions.  The concept of the Fermi surface approximation is discussed in detail in \cite{Alford:2018lhf} in the context of the Urca process.

The mean free path of an axion with energy $\omega$ due to (inverse) axion bremsstrahlung has been calculated in the Fermi surface approximation in \cite{Ishizuka:1989ts,Iwamoto:1992jp}, where they find 
\begin{equation}
\lambda_{FS}^{-1} = \frac{1}{18\pi^5}\frac{f^4G_{an}^2m_n^4}{m_{\pi}^4}p_{Fn}F(y)\frac{\omega^2+4\pi^2T^2}{1-e^{-\omega/T}},\label{eq:lambdaFS}
\end{equation}
where
\begin{equation}
F(y) = 4 - \frac{1}{1+y^2} + \frac{2y^2}{\sqrt{1+2y^2}}\arctan{\left(\frac{1}{\sqrt{1+2y^2}}\right)}-5y\arcsin{\left(\frac{1}{\sqrt{1+y^2}}\right)},
\label{eq:Fofy}
\end{equation}
with $y=m_{\pi}/(2p_{Fn})$.  The derivation of this formula is sketched in Appendix \ref{sec:mfp_FS_calculation}.  The axion MFP in the Fermi surface approximation is plotted in dotted lines in Fig.~\ref{fig:axionMFP}.  We see from Fig.~\ref{fig:axionMFP} that in conditions that are not strongly degenerate, the Fermi surface approximation significantly underestimates the mean free path compared to the constant-matrix-element approximation.

The virtue of the Fermi surface approximation is that it allows the 15 dimensional phase space integral to be done analytically.  However, during the course of the calculation, the lower endpoint of the integration over neutron energy (which comes from converting the phase space integral into spherical coordinates and then turning the momentum magnitude integral to an integral over energy - see Appendix \ref{sec:mfp_FS_calculation}) is extended to minus infinity.  Extending the integration bounds this way is valid in degenerate nuclear matter, because it adds only an exponentially small term \cite{Shapiro:1983du}.  However, as temperature increases, the extension of the integral gives rise to a sizeable error.  We propose here an improved Fermi surface approximation calculation which keeps the energy bounded by $U_n + m_* < E_n < \infty$.  From there it is possible to write the axion mean free path as a one dimensional integral
\begin{equation}
    \lambda^{-1}=\frac{4}{3\pi^5}\frac{f^4G_{an}^2m_n^4}{\omega m_{\pi}^4}p_{Fn}F(y)T^3K_1(\hat{y},\omega/T),\label{eq:lambda_new}
\end{equation}
where 
\begin{align}
    &K_1(\hat{y},\omega/T) \equiv \int_{-2\hat{y}}^{\infty}\mathop{du} \frac{1}{(1-e^u)(1-e^{-u-\omega/T})}\label{eq:K1}\\
    &\times\ln{\left\{\frac{\cosh{\left[(u+\hat{y}+\omega/T)/2\right]}}{\cosh{(\hat{y}/2)}}\right\}}\ln{\left\{\frac{\cosh{(\hat{y}/2)}}{\cosh{\left[(u+\hat{y})/2\right]}}\right\}},\nonumber
\end{align} and
$\hat{y}=(\mu_n^*-m_*)/T$.  The degeneracy parameter $\hat{y}$ is just the logarithm of the neutron fugacity $\hat{y}=\ln{z_n}$.  This expression for the mean free path follows the Fermi surface approximation, but better treats the lower endpoint of integration over the neutron energies, where the traditional treatment \cite{Shapiro:1983du} becomes increasingly poor.  The details are further explained in Appendix \ref{sec:mfp_FS_calculation} and \ref{sec:curve_fits}, where we also include a curve-fit of $K_1(\hat{y},\omega/T)$ which is valid as long as the matter is relatively degenerate ($z_n > 1$).  Our expression Eq.~(\ref{eq:lambda_new}) of course matches the Fermi surface approximation (\ref{eq:lambdaFS}) in the degenerate limit $\hat{y}\rightarrow \infty$.  

We emphasize that this proposed expression improves the behavior of the FS approximation in semi-degenerate conditions, but is definitely not valid for non-degenerate conditions.  After all, this approximation still assumes that only neutrons on their Fermi surface participate in the process.
 
%%%%%%%%%%%%%%%%%%%%%%%%%%%%%%%%%%%%%%%%%%%%%%%%%%%%%%%%%%%%%%%
\subsection{MFP dependence on the axion-neutron coupling}
\label{sec:axion_MFP_vs_G}
\begin{figure*}[t!]
\begin{minipage}[t]{0.5\linewidth}
\includegraphics[width=.95\linewidth]{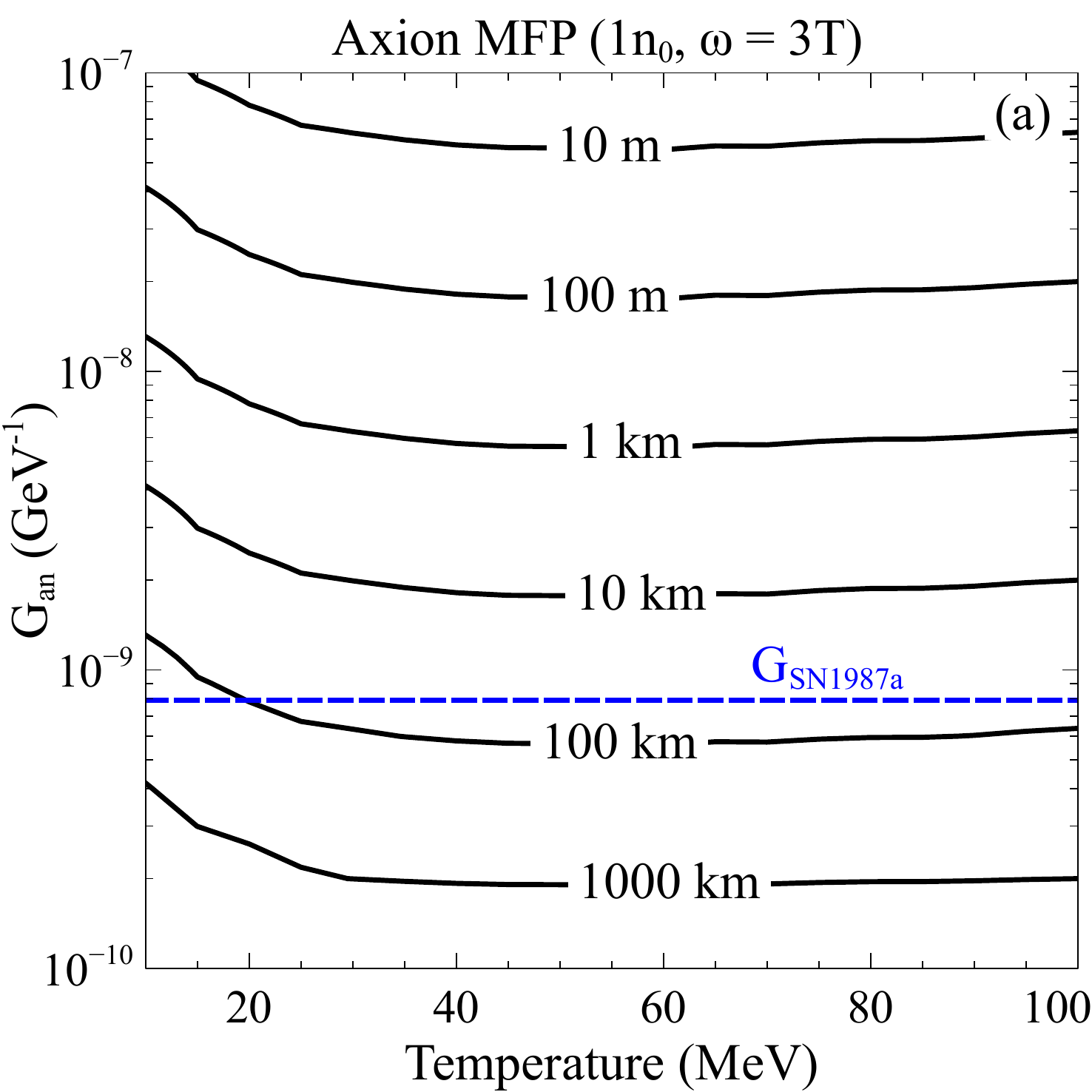}
\end{minipage}\hfill%
\begin{minipage}[t]{0.5\linewidth}
\includegraphics[width=.95\linewidth]{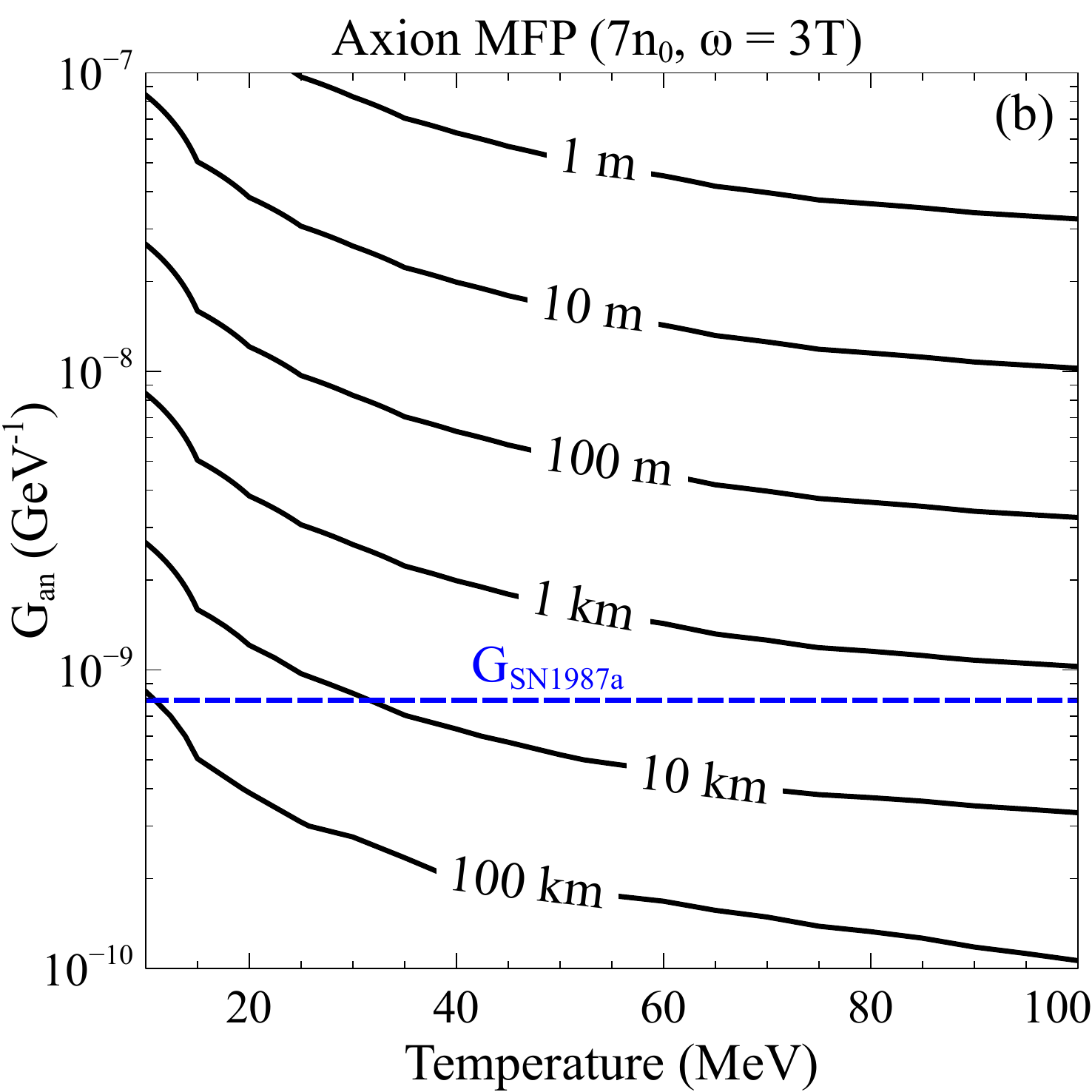}
\end{minipage}%
\caption{Axion mean free path (for axions with energy $\omega = 3T$) via $n+n+a\rightarrow n+n$ for densities of $1n_0$ (a) and $7n_0$ (b).  All couplings larger than the dotted blue line may be disallowed by the observation of SN1987A.  Thus, axions likely free-stream through neutron star mergers as their mean free path is above several kilometers regardless of the density.}
\label{fig:axionMFP_vs_T}
\end{figure*}

It is clear that if the axion-neutron coupling is less than or equal to the maximum value SN1987A will allow, all thermodynamic conditions encountered in mergers will fail to trap axions.  However, if the axion-neutron coupling was larger, the hotter, denser regions of the parameter space depicted in Fig.~\ref{fig:axionMFP} would begin to trap axions.  We depict this observation in Fig.~\ref{fig:axionMFP_vs_T}, using the constant-matrix-element approximation for the axion mean free path (and setting $\beta=0$ in the constant matrix element, for simplicity.)  We see that if the SN1987A bound is robust, for allowed values of the coupling, the axion mean free path is always comparable to or much bigger than the size of a neutron star and thus axions would free-stream from the star or merger.  However, if the SN1987A bounds are flexible due to uncertainties in modeling supernovae as well as uncertainties in the nuclear EoS (as is suggested by \cite{Bar:2019ifz,Fischer:2016cyd}, see also \cite{Giannotti:2017hny} for useful commentary), in the highest density and temperature conditions possibly reached in mergers, axions would be trapped for couplings just 4 times the SN1987 bound.  However, we note that there are many other independent proposed upper bounds on the axion-neutron coupling, coming from neutron star observations \cite{Berenji:2016jji,Sedrakian:2018kdm,Sedrakian:2015krq,Paul:2018msp,Beznogov:2018fda,Lloyd:2019rxg}, all within about an order of magnitude in either direction of the SN1987A bound.

%%%%%%%%%%%%%%%%%%%%%%%%%%%%%%%%%%%%%%
\section{Axion-transparent matter}
\label{sec:axion_transp_matter}
In the axion-transparent regime where the axion mean free path is comparable to or larger than the system size, an axion created inside the merger by neutron bremsstrahlung would escape, cooling down the merger.  This is analogous to neutrino cooling \cite{Yakovlev:2004iq,Potekhin:2015qsa}, which occurs in nuclear matter at temperatures below 5 or 10 MeV, which is the regime where it is transparent to neutrinos \cite{Alford:2018lhf,Roberts:2016mwj,Haensel:1987zz,1979ApJ...230..859S,Sawyer:1975js}.  We study regions of the merger above temperatures of 10 MeV, where neutrinos are trapped (and thus only cool via the relatively slow diffusion process \cite{1982ApJ...253..816G}) but axion emission could serve as an unexpected cooling mechanism.

We calculate the temperature of a fluid element radiating axions as a function of time, and in particular, find the characteristic cooling time at which the temperature has halved.  We can write a differential equation for the temperature $T$ as a function of time
\begin{equation}
    \frac{\mathop{dT}}{\mathop{dt}} = -\frac{Q}{c_V},\label{eq:dTdt}
\end{equation}
where $Q=\mathop{d\varepsilon}/\mathop{dt}$ is the axion emissivity (the energy emitted in axions per volume per time) and $c_V = \mathop{d\varepsilon}/\mathop{dT}$ is the specific heat of the nuclear matter per unit volume.  The specific heat is dominated by neutrons, as they have the largest particle fraction in the neutron star and thus the most possible low-energy excitations \cite{Alford:2017rxf}.  The specific heat\footnote{Technically, we should not use the specific heat for a degenerate Fermi gas as we do here, but instead for a Fermi gas at arbitrary degeneracy.  However, for the thermodynamic conditions encountered here, these differ by at most 14\%, the greatest difference occurring at large temperature.  Extending the traditional calculation \cite{1994ARep...38..247L} to the case of neutrons described by a relativistic mean field theory, the specific heat of a neutron gas of arbitrary degeneracy is $c_V = 2\int\frac{\mathop{d^3p}}{(2\pi)^3}(E^*-\mu_n^*)\frac{\mathop{d(f(E^*-\mu_n^*))}}{\mathop{dT}} = \frac{T^3}{4\pi^2}\int_{-\hat{y}_n}^{\infty}\mathop{dx}\frac{x^2(x+(\mu_n^*/T))\sqrt{(x+(\mu_n^*/T))^2-(m_*/T)^2}}{\cosh{(x/2)}^2}$, where $\hat{y}_n\equiv (\mu_n^*-m_*)/T$ and we have assumed that $m_*$ and $\mu_n^*$ do not depend on temperature.} is given by \cite{1994ARep...38..247L}
\begin{equation}
    c_V \approx \frac{1}{3}m_L p_{Fn}T,\label{eq:cV_fermion}
\end{equation}
where $m_L$ is the Landau effective mass of the neutron, which is related to its density of states at the Fermi surface \cite{Maslov:2015wba,Li:2018lpy}
\begin{equation}
    m_L = \frac{p_{Fn}}{(\mathop{dE}/\mathop{dp})\vert_{p_{Fn}}}=\sqrt{p_{Fn}^2+m_*^2}.\label{eq:Landau}
\end{equation}
The axion emissivity is given by the phase space integral \cite{Brinkmann:1988vi}
\begin{align}
&Q = \int \frac{\mathop{d^3p}_1}{\left(2\pi\right)^3}\frac{\mathop{d^3p}_2}{\left(2\pi\right)^3}\frac{\mathop{d^3p}_3}{\left(2\pi\right)^3}\frac{\mathop{d^3p}_4}{\left(2\pi\right)^3}\frac{\mathop{d^3\omega}}{\left(2\pi\right)^3}\frac{S\sum \vert \mathcal{M}\vert^2}{2^5 E_1^*E_2^*E_3^*E_4^*\omega}\omega\label{eq:emissivity_integral}\\
&\times\left(2\pi\right)^4\delta^4(p_1+p_2-p_3-p_4-\omega)f_1f_2\left(1-f_3\right)\left(1-f_4\right).\nonumber
\end{align}
In the rest of this section, we will discuss approximations of this axion emissivity phase space integral (just as we did for the axion MFP) and then we use our results to calculate the cooling time due to axion emission.
%%%%%%%%%%%%%%%%%%%%%%%%%%%%%%%%%%%%%%%%%%%%
\subsection{Relativistic, arbitrary degeneracy}
\label{sec:axion_Q_rel}
Like the calculation of the axion MFP, the axion emissivity involves an integration over phase space and we will make the same set of approximations we made for the MFP in Sec.~\ref{sec:axion_mfp_rel}.  Thus, assuming a momentum-independent matrix element Eq.~(\ref{eq:cst_matrix}) and using relativistic dispersion relation Eq.~(\ref{eq:En}) for the neutrons, we do the phase space integral [Eq.~(\ref{eq:emissivity_integral})] and find that the axion emissivity can be reduced to a six-dimensional integral to be done numerically.  The expression [Eq.~(\ref{eq:exact_emissivity})] and its derivation are given in Appendix \ref{sec:Q_rel_PS}.  That expression is valid for neutrons of arbitrary degeneracy.

In Appendix \ref{sec:Q_nonrel_PS}, we also present the phase space integral of the axion emissivity while assuming non-relativistic neutrons.  In this case, the momentum-dependence of the matrix element can be retained, and it is left inside the integral.  This calculation has been done before in the literature, but, as with the axion MFP integration, we present a version of the calculation well-adapted to a relativistic mean field theory.
%%%%%%%%%%%%%%%%%%%%%%%%%%%%%%%%%%%%%%%%%%%%%%%%%%%%%%%%%%%
\subsection{Fermi surface approximation (degenerate neutrons)}
\label{sec:axion_Q_FS}
As in Sec.~\ref{sec:axion_mfp_FS}, the Fermi surface approximation can be applied to the axion emissivity if the neutrons are strongly degenerate, as in that case only neutrons near the Fermi surface will participate in the bremsstrahlung process.  The calculation of the axion emissivity in this regime was done first by Iwamoto \cite{PhysRevLett.53.1198}, and extended by \cite{Iwamoto:1992jp, Stoica:2009zh}.  The Fermi surface approximation for the axion emissivity is
\begin{equation}
    Q_{FS} = \frac{31}{2835\pi}\frac{f^4 G_{an}^2m_n^4}{m_{\pi}^4}p_{Fn}F(y)T^6,\label{eq:Q_FS}
\end{equation}
where $F(y)$ is given in Eq.~(\ref{eq:Fofy}).  The derivation of this formula is sketched in Appendix \ref{sec:Q_FS_derivation}.  

As with the mean free path, discussed in Sec.~\ref{sec:axion_mfp_FS}, the Fermi surface approximation of the emissivity extends the lower endpoint of integration of neutron energy down to $-\infty$.  We propose an improvement to the FS approximation which keeps the neutron energy bounded by $m_*+U_n<E_n<\infty$, at the cost of having an emissivity expression in terms of a two-dimensional integral instead of an analytic expression like Eq.~(\ref{eq:Q_FS}).  The axion emissivity in the improved FS approximation is
\begin{equation}
    Q = \frac{2}{3\pi^7}\frac{f^4G_{an}^2m_n^4}{m_{\pi}^4}p_{Fn}F(y)T^6K_2(\hat{y})\label{eq:Q_new}
\end{equation}
where
\begin{align}
    K_2(\hat{y}) &= \int_{-2\hat{y}}^{\infty}\mathop{du}\frac{1}{1-e^u}\ln{\left\{\frac{\cosh{(\hat{y}/2)}}{\cosh{\left[(u+\hat{y})/2\right]}}\right\}}\label{K2}\\
    &\times \int_0^{u+2\hat{y}}\mathop{dw}\frac{w^2}{1-e^{w-u}}\ln{\left\{\frac{\cosh{\left[(u+\hat{y}-w)/2\right]}}{\cosh{(\hat{y}/2)}}\right\}}.\nonumber
\end{align}
%%%%%%%%%%%%%%%%%%%%%%%%%%%%%%%%%%%%%%%%%%
\subsection{Radiative cooling time dependence on axion-neutron coupling}
\label{sec:axion_cooling}
In the Fermi surface approximation, the differential equation (\ref{eq:dTdt}) for $T(t)$ can be solved exactly (assuming the neutron Landau effective mass does not depend on temperature, which is a reasonable approximation) and we find a fluid element that starts at temperature $T_0$ cools according to
\begin{equation}
    T(t)^{-4}=T_0^{-4}+\frac{124}{945\pi} \frac{f^4G_{an}^2m_n^4F(y)}{m_Lm_{\pi}^4}t,
\end{equation}
and thus has cooling time (to reach half of its initial temperature)
\begin{equation}
    \tau_{FS,1/2} \approx 12 \text{s} \frac{\left(\dfrac{m_L}{0.8m_n}\right)}{\left(\dfrac{G_{an}}{G_{SN1987A}}\right)^{\!2}F(y)\left(\dfrac{T_0}{10\text{ MeV }}\right)^{\!4}} .\label{eq:t_cool_FS}
\end{equation}

\begin{figure}[tbp]
\centering
\includegraphics[width=.45\textwidth]{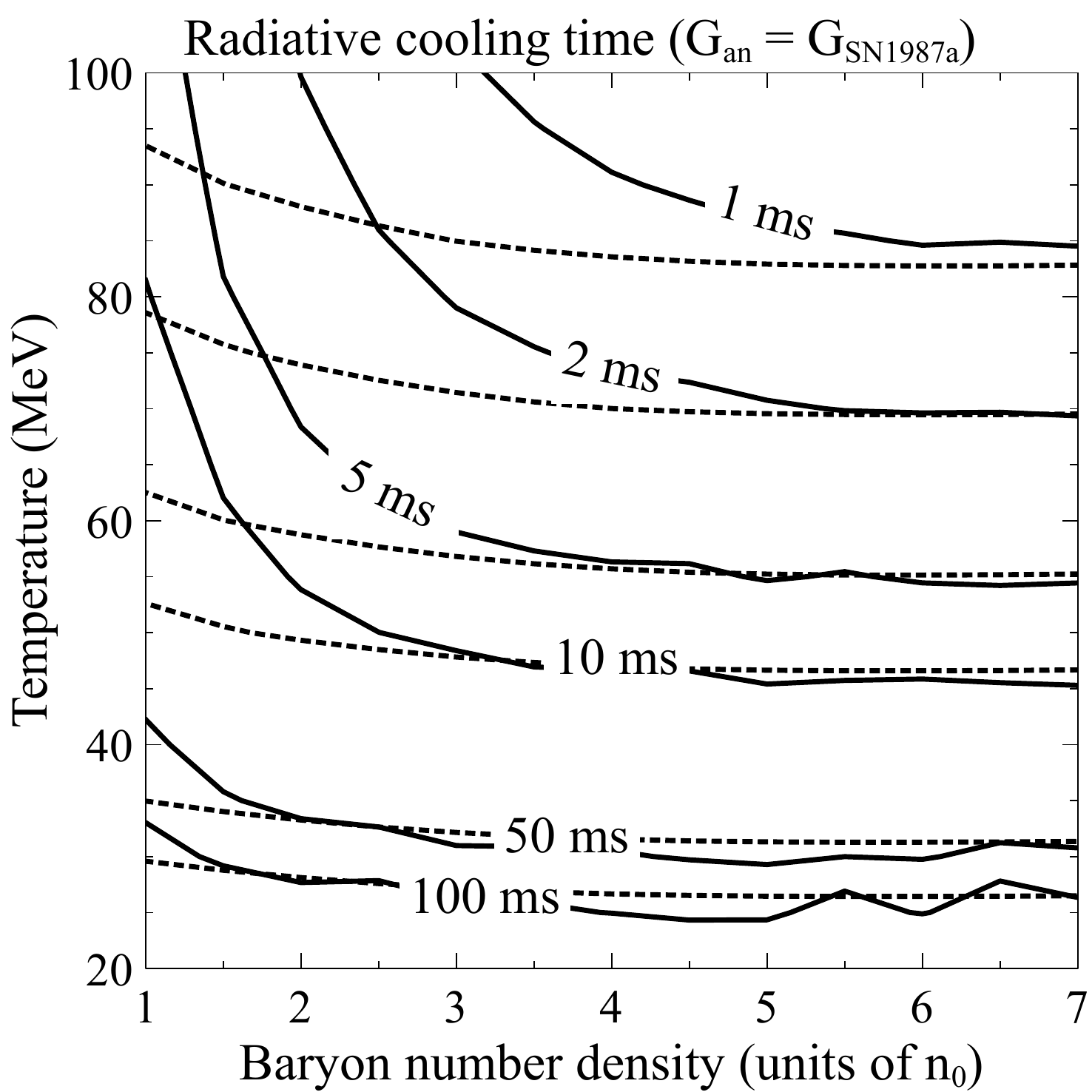}
\caption{Cooling time of nuclear matter due to radiating axions.  Solid lines use the constant-matrix-element approximation of the emissivity while dotted lines use the FS approximation.  The axion-neutron coupling is chosen to be the bound set by SN1987A.}
\label{fig:radiative_cooling_dens_temp}
\end{figure}

While there is no analytic result for the characteristic cooling time with our new expression for the emissivity Eq.~(\ref{eq:exact_emissivity}), we can solve the differential equation numerically and plot the characteristic cooling time as a function of density and temperature for a particular choice of axion-neutron coupling constant.  In Fig.~\ref{fig:radiative_cooling_dens_temp}, we plot the radiative cooling time due to axion emission, choosing the coupling constant to be the maximum value allowed by SN1987A.  The solid contours use the constant-matrix-element phase space integral for the emissivity (with $\beta=0$), while the dotted contours use the FS approximation.  We see that hotter and denser regions cool faster, because they emit axions at a higher rate.  The solid and dashed contours agree where the nuclear matter is strongly degenerate, which occurs at high density and low temperature.  This plot indicates that within the constraints set by SN1987A, axions can cool fluid elements in timescales relevant to neutron star mergers.  The treatment of the axion emissivity via the full phase space integration limits (compared to the FS approximation) the range of densities for which fast cooling can occur.  In particular, hot nuclear matter near saturation density has a significantly longer cooling time than predicted by the FS approximation.

\begin{figure*}[t!]
\begin{minipage}[t]{0.5\linewidth}
\includegraphics[width=.95\linewidth]{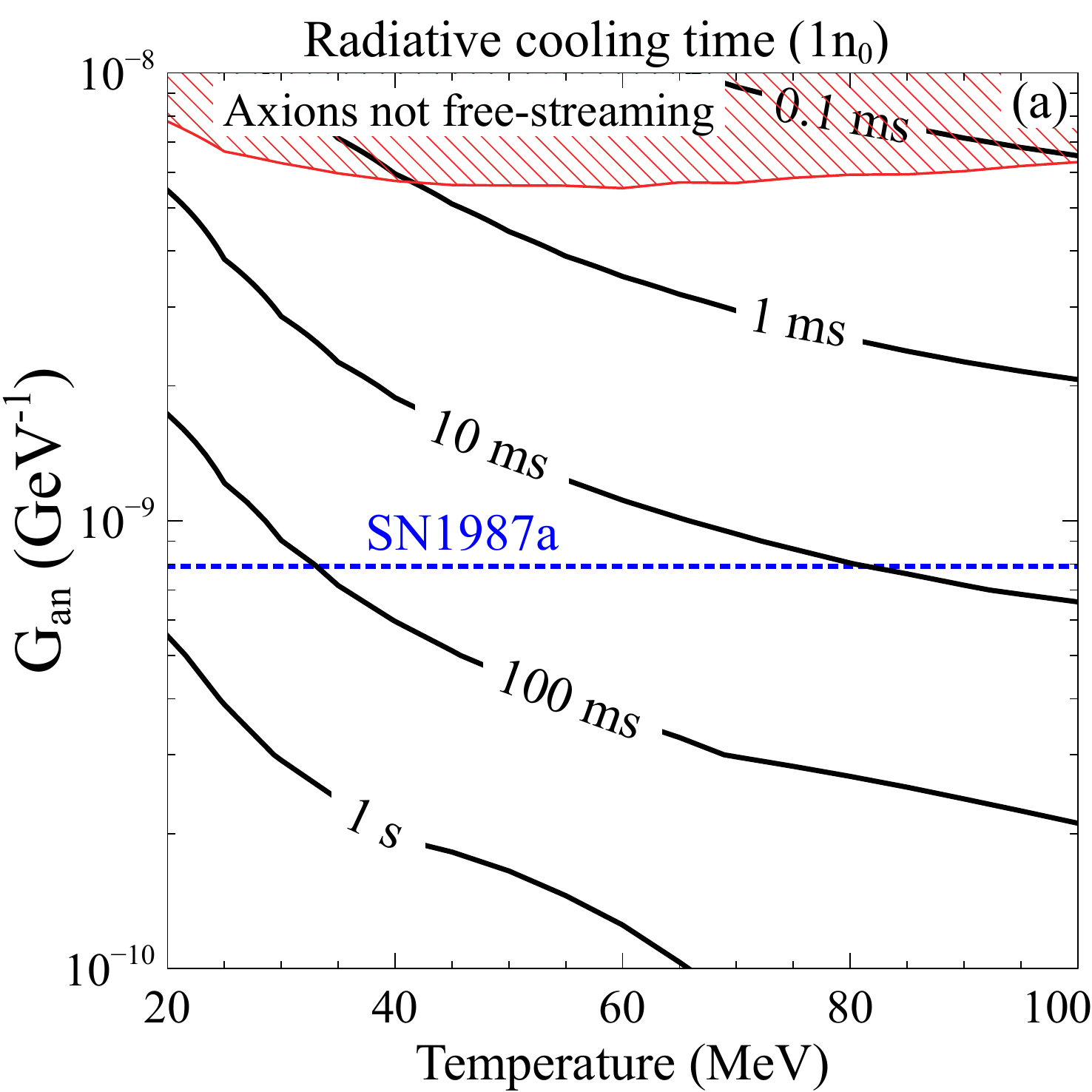}
\end{minipage}\hfill%
\begin{minipage}[t]{0.5\linewidth}
\includegraphics[width=.95\linewidth]{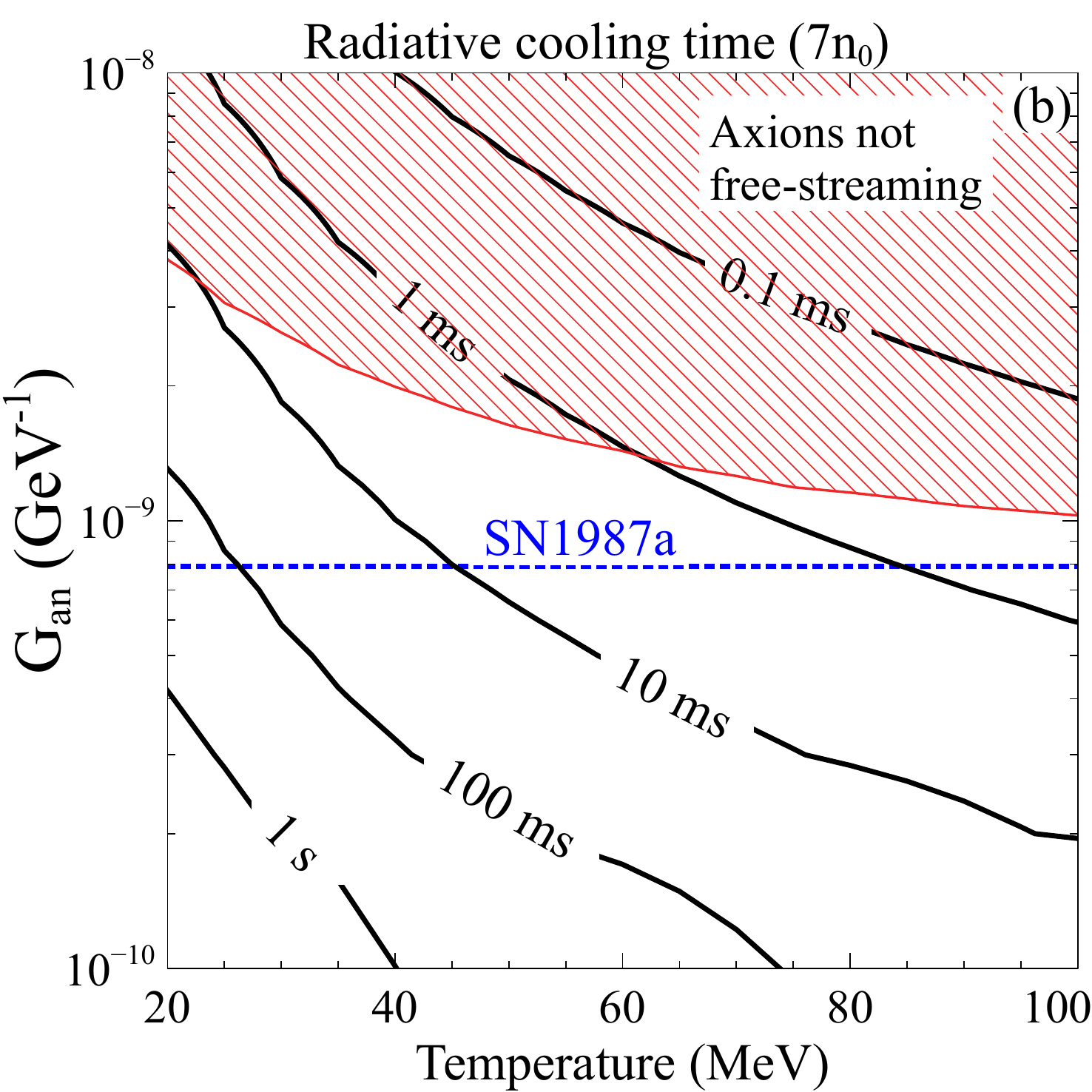}
\end{minipage}%
\caption{Radiative cooling time due to axion emission at densities of $1n_0$ (a) and $7n_0$ (b).  All couplings stronger than the dotted blue line are disallowed by the observation of SN1987A.  Considering couplings compatible with SN1987A, lower density regions cool somewhat slowly compared to merger timescales, but higher density regions could cool on timescales relevant for mergers.}
\label{fig:rad_cool_time_vs_G}
\end{figure*}
In Fig.~\ref{fig:rad_cool_time_vs_G}, we plot the radiative cooling time at two different densities, as a function of temperature and axion-neutron coupling. Radiative cooling is only relevant for thermodynamic conditions where the axions have a mean free path longer than a few kilometers.  At $1n_0$, it is difficult to get substantial cooling on merger timescales, but at high densities ($7n_0$), cooling times can be under 10 ms, and definitely under 100 ms for a wide range of temperatures.  

It is worth comparing our calculation of the axion emissivity (Eq.~\ref{eq:exact_emissivity}) to the expression used by Dietrich \& Clough \cite{Dietrich:2019shr}, which was adapted from Brinkmann \& Turner \cite{Brinkmann:1988vi}.  For a particular EoS and a fixed value of the axion-neutron coupling, the expression for emissivity derived in this paper can be as much as 3 times larger than the Dietrich \& Clough result.  The difference at low temperatures is small, but grows at high temperature.  This means that our predicted cooling times for the hottest fluid elements in the merger will be up to a few times shorter than those in the Dietrich \& Clough result, for the same EoS and axion-neutron coupling.

When comparing different nuclear equations of state, our results for the axion emissivity change by under 70\% for the few EoSs that we tried. This difference seems to come from the different predictions for the neutron fugacity as a function of baryon density and temperature, which arises from different values for the meson mean fields.  In this work we consider only nuclear matter consisting of neutrons, protons, electrons, and neutrinos.  If we were to consider exotic phases, like quark matter or hyperons, the axion emissivity might be quite different from what we predict here.  The same is of course true for the axion mean free path calculation in the previous section.
%%%%%%%%%%%%%%%%%%%%%%%%%%%%%%%%%%%%%%%%%%%%%%%%%%%%%%%%%%%%%%%%%%%
\section{Axion-trapped matter}
\label{sec:axion_trapped}
Based on our results in Fig.~\ref{fig:axionMFP}, we do not expect axions to be trapped in any part of a neutron star merger.  However, we present the following analysis of trapped axions for completeness, but also as an example of the contribution to thermal equilibration of the interior of a neutron star due to a boson that interacts with neutrons. 

If the mean free path of axions is much less than the system size, then the axions form a Bose gas inside the neutron star merger.  In this situation, the axions could transport energy around the star, smoothing out temperature gradients, much like neutrinos do when they are trapped.  We calculate the timescale of thermal equilibration for a fluid element to transfer heat to its neighboring fluid elements.  
From \cite{Alford:2017rxf}, a hot spot of volume $z^3$ in the merger has extra thermal energy (compared to its neighbors) $E_{\text{th}} =  c_V z^3 \Delta T$, and it conducts that energy away through its boundaries at rate $W_{\text{th}} = \kappa (\Delta T/z)6z^2$.  Thus, the timescale for heat conduction is
\begin{equation}
    \tau_{\kappa} = E_{th}/W_{th} = \frac{c_V z^2}{6\kappa}.\label{eq:t_conduct_general}
\end{equation}
As discussed in Sec.~\ref{sec:axion_transp_matter}, the specific heat $c_V$ is dominated by the neutrons, which have specific heat $c_V = (1/3)m_L p_{Fn} T$.  The thermal conductivity is the sum of the contributions $\kappa_i = (1/3)c_{V_i}v_i\lambda_i$ from each particle species.  The particles with both high density and long mean free path (but still less than the system size) will dominate the thermal conductivity.  We consider here only stars with temperatures above 10 MeV, which mean that neutrinos will be trapped and would traditionally dominate the thermal conductivity \cite{Alford:2017rxf}.  However, if the star also traps axions, then axions could take over the role of energy transportation.

The neutrinos have \cite{1982ApJ...253..816G} thermal conductivity $\kappa_{\nu} \approx n_{\nu}^{2/3}/(3G_F^2m_L^2n_e^{1/3}T)$, which implies that neutrinos re-establish thermal equilibrium between nearby fluid elements in time \cite{Alford:2017rxf}
\begin{equation}
    \tau_{\nu} = 700 \text{ ms}\left(\frac{0.1}{x_p}\right)^{1/3}\left(\frac{m_L}{0.8m_n}\right)^3\left(\frac{\mu_e}{2\mu_{\nu}}\right)^2\left(\frac{z}{1 \text{ km}}\right)^2\left(\frac{T}{10 \text{ MeV}}\right)^2.\label{eq:conduct_nu}
\end{equation}
When two species contribute to thermal equilibration, their individual timescales add according to $\tau_{\kappa}^{-1} = \tau_{\kappa_a}^{-1}+\tau_{\kappa_{\nu}}^{-1}$.  As we will see, the equilibration timescale due to axions is much shorter (for the range of couplings we consider) than the timescale due to neutrinos, and so for the rest of this discussion we assume axions are the only species contribution to thermal equilibration and thus
\begin{equation}
    \tau_{\kappa} \approx \frac{c_{V_n} z^2}{6\kappa_a}.
\end{equation}

As the axions, when trapped, are a free Bose gas with zero chemical potential (they are equilibrated by the reaction $n+n+a\leftrightarrow n+n$), they have energy density $\varepsilon = (\pi^2/30)T^4$ and thus specific heat per unit volume $c_V = (2\pi^2/15)T^3$, and so their thermal conductivity is $\kappa_a = (2\pi^2/45)T^3\lambda_a$.  The axion conduction timescale is
\begin{equation}
    \tau_{a} = \frac{5}{4\pi^2}\frac{m_Lp_{Fn}z^2}{T^2\lambda},
    \label{eq:conductive_time_general}
\end{equation}
which in the Fermi surface approximation for the MFP, reduces to
\begin{equation}
    \tau_{\kappa,a,FS} \approx 2.757 f^4G_{an}^2m_Lp_{Fn}^2F(y)z^2.
    \label{eq:t_conduct_FS}
\end{equation}
Eq.~(\ref{eq:conductive_time_general}) indicates that the longer the axion mean free path, the shorter the timescale for thermal equilibration.  However, if the mean free path is too long, then there is no heat conduction due to axions.

To get an estimate for the thermal equilibration timescale, we consider the situation where a neutron star merger has only gradual temperature gradients, occurring on at least the 1 km scale.  For example, the hot spherical shell observed in many simulations \cite{Hanauske:2016gia, Perego:2019adq,Hanauske:2019qgs,Hanauske:2019czl,Hanauske:2019vsz} is 1-2 km thick.  If this is the case, then neutrino conduction has no effect on a neutron star merger, as heat conduction via neutrinos occurs on timescales greater than one second.

\begin{figure*}[t!]
\begin{minipage}[t]{0.5\linewidth}
\includegraphics[width=.95\linewidth]{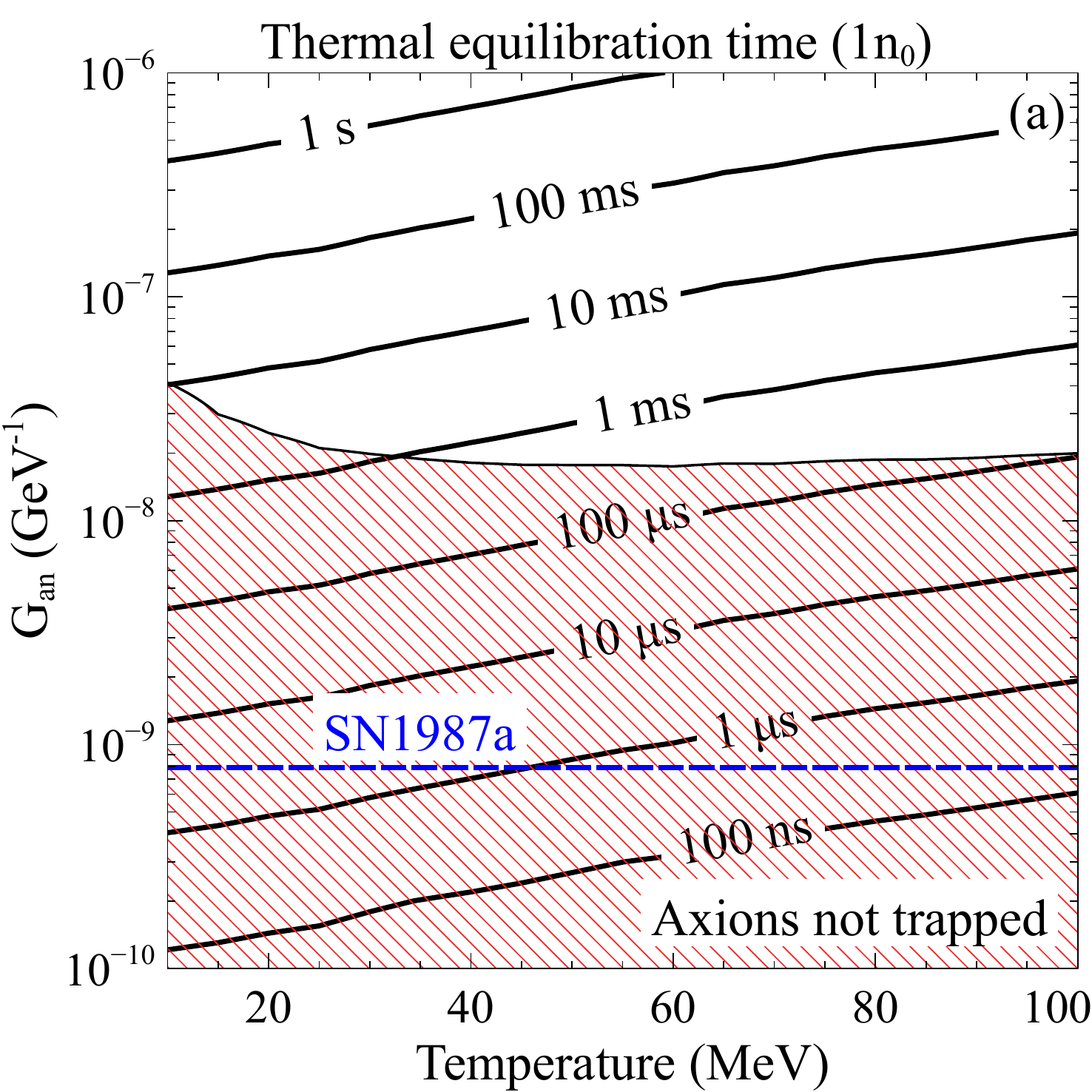}
\end{minipage}\hfill%
\begin{minipage}[t]{0.5\linewidth}
\includegraphics[width=.95\linewidth]{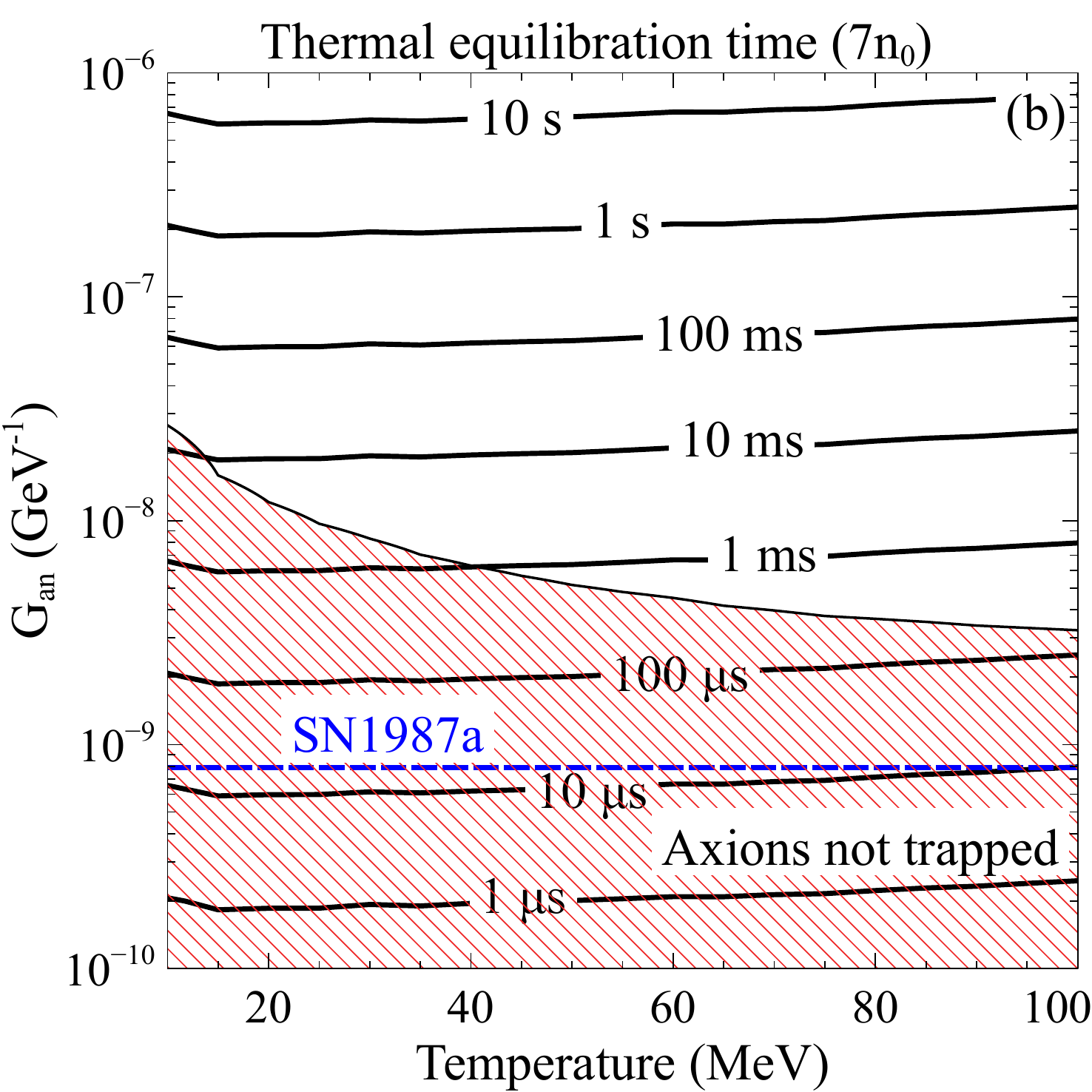}
\end{minipage}%
\caption{Thermal equilibration time due to axion conduction at a density of $1n_0$ (a) and $7n_0$ (b).  All couplings stronger than the dotted blue line are disallowed by the observation of SN1987A.  The red shaded region corresponds to couplings at which axions are not trapped, and so there is no axion conduction between neighboring fluid elements.}
\label{fig:conductivecooling}
\end{figure*}

According to Fig.~\ref{fig:conductivecooling}, at low axion-neutron couplings, the thermal equilibration due to axion conduction would occur very quickly because of the long axion mean free path.  However, the mean free path is too long - bigger than the size of a neutron star - and so neighboring fluid elements are unable to transfer heat to each other via axions.  As the coupling gets strong enough to trap axions, the thermal equilibration time is still fast enough to be relevant.  However, this only occurs at axion coupling constants more than an order of magnitude higher than the limit set by SN1987A, which is why thermal equilibration of neutron star mergers due to axions is unlikely.  
%%%%%%%%%%%%%%%%%%%%%%%%%%%%%%%%%%%%%%%%%%%%%
\section{Conclusion}
\label{sec:conclusion}
We have analyzed the impact of axions on neutron stars mergers in the case that they are trapped and in the case that they are free-streaming.  As part of this effort, we calculated the axion mean free path and emissivity due to the neutron bremsstrahlung process $n+n\leftrightarrow n+n+a$.  In contrast to previous calculations, we integrated over the entire phase space while using a relativistic treatment of the neutrons (although assuming the matrix element was momentum-independent).  In particular, we used a relativistic mean field theory to describe the nucleons, which means that we took into account the precipitous decrease in the (Dirac) effective mass of the nucleons as density increases above nuclear saturation density.

With our calculation of the axion mean free path, we were able to, for a given value of the axion-neutron coupling, divide the thermodynamic parameter space $\{n_B,T\}$ into regions where axions are trapped and where they are free-streaming (and of course, a difficult-to-treat region in between where they are neither).  We find that for axion-neutron couplings allowed by SN1987A, axions have a long mean free path in all thermodynamic conditions encountered in mergers (see Fig.~\ref{fig:axionMFP_vs_T}), and thus we expect them to free-stream from a neutron star merger.

We examine the time it would take for a fluid element of nuclear matter in a merger to cool to one half of its current temperature by radiating axions.  The result is depicted in Figs.~\ref{fig:radiative_cooling_dens_temp} and \ref{fig:rad_cool_time_vs_G}.  For allowed values of the axion-neutron coupling, we find that the hottest fluid elements in a merger could cool in timescales less than 10 ms, which would have an impact on the dynamics of a neutron star merger, in particular, reducing the thermal pressure of the possible remnant \cite{Dietrich:2019shr} and changing the values of temperature-dependent transport properties like bulk viscosity \cite{Alford:2017rxf,Alford:2019kdw, Alford:2019qtm}.  However, the radiative cooling time increases dramatically as the axion-neutron coupling decreases.

Dietrich \& Clough \cite{Dietrich:2019shr} ran simulations of neutron star mergers which included the effects of cooling due to axion emission.  They concluded that even though axion emission reduces the temperature and, consequently, the lifetime of the remnant, as well as slightly increases the density of the core, these effects would not modify the gravitational wave signal or the amount of ejected material enough to be observable in the near future.  In this paper, we find that the axion cooling timescale for the hottest fluid elements in the merger could be reduced by a factor of a few, for a fixed value of the axion-neutron coupling.  In addition, we point out that the changes to the density and temperature of the remnant resulting from the emission of axions could trigger phase transitions in the nuclear matter, or change the transport processes acting in the remnant, which could themselves have observational consequences.

We also consider the possibility of axion trapping, where axion  diffusion might serve to thermally equilibrate the interior of a neutron star or merger.  The timescales of thermal equilibration are given in Fig.~\ref{fig:conductivecooling}.  The SN1987A bound rules out axion trapping, but we present the analysis of thermal equilibration as it is potentially useful for the study of thermal transport due to future proposed bosons in neutron stars.  

Finally, in the appendix we present in detail our calculations of the axion mean free path and emissivity.  We first present the phase space integrals assuming relativistic neutrons of arbitrary degeneracy but a momentum-independent matrix element.  Then we present the phase space integrals assuming non-relativistic neutrons of arbitrary degeneracy and a momentum-dependent matrix element.  Finally, we present the Fermi surface approximation, which is valid for strongly degenerate neutrons, and we propose a better treatment of the energy integration in that approximation which extends the usefulness of the Fermi surface approximation to the semi-degenerate regime.  

In our calculations, we have neglected the interaction of protons and axions, which should further reduce the mean free path and increase the emissivity \cite{Brinkmann:1988vi,Paul:2018msp,Iwamoto:1992jp,Keil:1996ju,PhysRevD.52.1780}.  We have also used a very simplistic neutron-neutron interaction (one pion exchange), improvements to which have been discussed in \cite{Hanhart:2000ae,PhysRevD.52.1780,Keil:1996ju,Beznogov:2018fda,Bartl:2016iok}, which indicate that more sophisticated nuclear interactions decrease the emissivity and increase the axion mean free path at high densities.  In addition, the axion mass and the axion-nucleon coupling may themselves depend on density. A recent calculation \cite{Balkin:2020dsr} raises the possibility of a large enhancement of the coupling as the density rises, which would lead to shorter mean free paths at higher densities.

Future inclusion of axions in merger simulations should be done consistently.  The axion opacity should be calculated using the same nuclear equation of state used in the merger simulation itself \cite{Horowitz:2003yx,Fischer:2018kdt}, and the nucleon effective mass should be taken into account as discussed in Sec.~\ref{sec:axion_mfp_rel}.  This parallels the improvements of the neutrino transport in supernovae and merger simulations over the past couple of decades \cite{Ardevol-Pulpillo:2018btx,Perego:2014qda,Galeazzi:2013mia,Sekiguchi:2012uc,Rosswog:2003rv,1999JCoAM.109..281M}.

%%%%%%%%%%%%%%%%%%%%%%%%%%%%%%%%%%%%%%%%%%%
\section{Acknowledgments}
SPH acknowledges discussions with Francesc Ferrer and Alex Haber.  The work of MGA and SPH was partly supported by the U.S. Department of Energy, Office of Science, Office of Nuclear Physics, under Award No.~\#DE-FG02-05ER41375.  The work of JFF is supported by NSERC and FRQNT. KS is supported by  DOE Grant DE-SC0009956. He would like to thank KITP Santa Barbara for hospitality during part of the time that this work was completed. 

%%%%%%%%%%%%%%%%%%%%%%%%%%%%%%%%%
%%%%%%%%%%%%%%%%%%%%%%%%%%%%%
%%%%%%%%%%%%%%%%%%%%%%%%%%%%%%%%%
%%%%%%%%%%%%%%%%%%%%%%%%%%%%%%%%%%
\appendix
\section{Axion emissivity integrals}
Below, we detail a series of approximations for the axion emissivity, Eq.~(\ref{eq:emissivity_integral}).
\subsection{Relativistic, constant-matrix-element phase space integration}
\label{sec:Q_rel_PS}
The phase space integral Eq.~(\ref{eq:emissivity_integral}) can be done if we make the approximation of a momentum-independent matrix element [Eq.~(\ref{eq:cst_matrix})].  The resulting emissivity will be valid at arbitrary neutron degeneracy and arbitrary degree of relativistic nature of the neutrons.  
In the constant-matrix element approximation the axion emissivity is
\begin{align}
    Q &= \left(1-\frac{\beta}{3}\right)\frac{f^4m_n^4G_{an}^2}{256\pi^{11}m_{\pi}^4}\left(1+\frac{m_{\pi}^2}{k_{\text{typ}}^2}\right)^{-2}\\
    &\times \int \mathop{d^3p_1}\mathop{d^3p_2}\mathop{d^3p_3}\mathop{d^3p_4}\mathop{d^3\omega}\delta^4(p_1+p_2-p_3-p_4-\omega)\frac{f_1f_2(1-f_3)(1-f_4)}{E_1^*E_2^*E_3^*E_4^*}\nonumber.
\end{align}

The zeroth component of the delta function can be written as $\delta(E_1^*+E_2^*-E_3^*-E_4^*-\omega)$ since the neutron mean fields $U_n$ cancel out.
This integral can be broken up into 2 subsystems $A$ and $B$ which exchange some 4-momentum $q$, which is integrated over (a similar approach was used by \cite{Kaminker:2016ayg} in a different context).  Thus,
\begin{equation}
    Q = \left(1-\frac{\beta}{3}\right)\frac{f^4m_n^4G_{an}^2}{256\pi^{11}m_{\pi}^4} \left(1+\frac{m_{\pi}^2}{k_{\text{typ}}^2}\right)^{-2}\int \mathop{d^4q} A(q_0,q) B(q_0,q)
\end{equation}
where 
\begin{equation}
    A(q_0,q) = \int \mathop{d^3p_1}\mathop{d^3p_2}\delta^4(p_1+p_2-q)\frac{f_1f_2}{E_1^*E_2^*}
\end{equation}
and
\begin{equation}
    B(q_0,q) = \int \mathop{d^3p_3}\mathop{d^3p_4}\mathop{d^3\omega}\delta^4(q-p_3-p_4-\omega)\frac{(1-f_3)(1-f_4)}{E_3^*E_4^*}.
\end{equation}
Then we split $A$ and $B$ up into subsystem with 4-momentum transfers $k$ and $l$
\begin{align}
    A(q_0,q) &= \int\mathop{d^4k}I_1(k_0,k)I_2(k_0,k)\label{eq:A_integral}\\
    B(q_0,q) &= \int\mathop{d^4l}I_3(l_0,l)I_4(l_0,l)\label{eq:B_integral}
\end{align}
where we define and compute
\begin{align}
    I_1 \equiv \int\mathop{d^3p_1}\delta^4(p_1+k)\frac{f_1}{E_1^*} &= \frac{\delta(k_0+\sqrt{\mathbf{k}^2+m_*^2})}{\sqrt{\mathbf{k}^2+m_*^2}(1+e^{(\sqrt{\mathbf{k}^2+m_*^2}-\mu_n^*)/T})}\\
    I_2 \equiv \int\mathop{d^3p_2}\delta^4(p_2-q-k)\frac{f_2}{E_2^*} &= \frac{\delta(\sqrt{m_*^2+(\mathbf{q}+\mathbf{k})^2}-q_0-k_0)}{\sqrt{m_*^2+(\mathbf{q}+\mathbf{k})^2}(1+e^{(\sqrt{m_*^2+(\mathbf{q}+\mathbf{k})^2}-\mu_n^*)/T})}\\
    I_3 \equiv \int\mathop{d^3p_3}\delta^4(q-p_3-l)\frac{(1-f_3)}{E_3^*} &= \frac{\delta(q_0-l_0-\sqrt{m_*^2+(\mathbf{q}-\mathbf{l})^2})}{\sqrt{m_*^2+(\mathbf{q}-\mathbf{l})^2}(1+e^{-(\sqrt{m_*^2+(\mathbf{q}-\mathbf{l})^2)}-\mu_n^*)/T})}\\
    I_4 \equiv \int\mathop{d^3p_4}\mathop{d^3\omega}\delta^4(l-p_4-\omega)\frac{(1-f_4)}{E_4^*} &= \frac{2\pi}{l}\int_{\omega_-}^{\omega_+}\mathop{d\omega}\frac{\omega}{1+e^{-(l_0-\omega-\mu_n^*)/T}}\theta(l_0^2-l^2-m_*^2),
\end{align}
where
\begin{align}
    \omega_- &= \frac{l_0^2-l^2-m_*^2}{2(l_0+l)}\\
    \omega_+ &= \frac{l_0^2-l^2-m_*^2}{2(l_0-l)}.
\end{align}
We can now calculate $A(q)$ and $B(q)$ using Eq.~(\ref{eq:A_integral}) and (\ref{eq:B_integral}).  We find
\begin{align}
    A(q_0,q) &= \frac{2\pi}{q}\int_0^{\sqrt{q_0^2-m_*^2}}\mathop{dk}\frac{k(q_0-\sqrt{k^2+m_*^2})}{\sqrt{k^2+m_*^2}\sqrt{k^2+m_*^2+q_0^2-2q_0\sqrt{k^2+m_*^2}}}\\
    &\times \frac{\theta(2kq-\vert q_0^2-q^2-2q_0\sqrt{k^2+m_*^2}\vert)}{(1+e^{(\sqrt{k^2+m_*^2}-\mu_n^*)/T})(1+e^{(\sqrt{k^2+m_*^2+q_0^2-2q_0\sqrt{k^2+m_*^2}}-\mu_n^*)/T})}\nonumber
\end{align}
and
\begin{align}
    B(q_0,q) = \frac{4\pi^2}{q}\int_m^{q_0}\mathop{dl_0}\int_0^{\sqrt{l_0^2-m_*^2}}\mathop{dl}\frac{\theta(2ql-\vert m_*^2+q^2+l^2-q_0^2-l_0^2+2q_0l_0\vert)}{1+e^{-(q_0-l_0-\mu_n^*)/T}}\times\int_{\omega_-}^{\omega_+}\mathop{d\omega}\frac{\omega}{1+e^{-(l_0-\omega-\mu_n^*)/T}}.
\end{align}
Thus, the full expression for the emissivity is
\begin{align}
    Q &= \left(1-\frac{\beta}{3}\right) \frac{f^4m_n^4G_{an}^2}{8\pi^7m_{\pi}^4}\left(1+\frac{m_{\pi}^2}{k_{\text{typ}}^2}\right)^{-2}\int_{m_*}^{\infty}\mathop{dq_0}\int_0^{\infty}\mathop{dq}\int_0^{\sqrt{q_0^2-m_*^2}}\mathop{dk}\int_{m_*}^{q_0}\mathop{dl_0}\int_0^{\sqrt{l_0^2-m_*^2}}\mathop{dl}\int_{\omega_-(l_0,l)}^{\omega_+(l_0,l)}\mathop{d\omega}\nonumber\\
    &\times k\omega(q_0-\sqrt{k^2+m_*^2})\frac{\theta(2kq-\vert q_0^2-q^2-2q_0\sqrt{k^2+m_*^2}\vert)\theta(2ql-\vert m_*^2+q^2+l^2-q_0^2-l_0^2+2q_0l_0\vert)}{\sqrt{k^2+m_*^2}\sqrt{k^2+m_*^2+q_0^2-2q_0\sqrt{k^2+m_*^2}}}    \label{eq:exact_emissivity}\\
    &\times \frac{1}{(1+e^{(\sqrt{k^2+m_*^2}-\mu_n^*)/T})(1+e^{(\sqrt{k^2+m_*^2+q_0^2-2q_0\sqrt{k^2+m_*^2}}-\mu_n^*)/T})(1+e^{-(q_0-l_0-\mu_n^*)/T})(1+e^{-(l_0-\omega-\mu_n^*)/T})}.\nonumber
\end{align}
This six-dimensional integral can be done numerically in Mathematica.
%%%%%%%%%%%%%%%%%%%%%%%%%%%%%%%%%%%%%%%%%%%%%%%%%%%%%%%%
\subsection{Non-relativistic phase space integration}
\label{sec:Q_nonrel_PS}
The axion emissivity [Eq.~(\ref{eq:emissivity_integral})] can be computed assuming non-relativistic neutrons.  In this case, it is possible to keep the full momentum-dependence of the matrix element [Eq.~(\ref{eq:matrix_element})].  However, the axion 3-momentum is neglected in the 3-momentum conserving delta function.  Calculations similar to this have been done in the literature, mostly for non-degenerate nucleons \cite{PhysRevD.52.1780,Giannotti:2005tn,Stoica:2009zh,Dent:2012mx,Turner:1987by}.  The full calculation of the axion emissivity with non-relativistic neutrons and at arbitrary degeneracy has been done recently by \cite{Carenza:2019pxu}.  Here we will apply this calculation to neutrons described by the NL$\rho$ EoS, that is, neutrons with dispersion relation given by Eq.~(\ref{eq:En}).  We will do the calculation for arbitrary degeneracy, using Fermi-Dirac factors instead of a Maxwell-Boltzmann distribution.  In the nonrelativistic approximation, $E-\mu_n = E^* - \mu_n^* \approx m^*+p^2/(2m^*)-\mu^* = p^2/(2m^*)-(\mu^*-m^*) \equiv p^2/(2m^*)-\hat{\mu}$, where $\hat{\mu} \equiv \mu^*-m^*$ is the non-relativistic definition of the chemical potential.  

Starting with Eq.~(\ref{eq:emissivity_integral}), we neglect the axion three-momentum in the three-dimensional delta function and we set the neutron energy $E^* = m^*$, in the factor $(2^5E_1E_2E_3E_4\omega)^{-1}$ as is conventional \cite{Dent:2012mx}, as it often simplifies the integration.  Converting the integral over the axion momentum to spherical coordinates and doing the trivial integral over the axion momentum solid angle, we obtain
\begin{align}
    Q_a &= \frac{1}{96\pi^{10}}\frac{f^4 m_n^4 G_{an}^2}{m_{\pi}^4m_*^3}\int \mathop{d^3p_1}\mathop{d^3p_2}\mathop{d^3p_3}\mathop{d^3p_4}\int_0^{\infty}\mathop{d\omega}\omega^2\delta(p_1^2+p_2^2-p_3^2-p_4^2-2m_*\omega)\delta^3(\mathbf{p}_1+\mathbf{p}_2-\mathbf{p}_3-\mathbf{p}_4)\nonumber\\
    &\times f_1f_2(1-f_3)(1-f_4)\left[ \frac{\mathbf{k}^4}{\left(\mathbf{k}^2+m_{\pi}^2\right)^2}+\frac{\mathbf{l}^4}{\left(\mathbf{l}^2+m_{\pi}^2\right)^2}+\frac{\mathbf{k}^2\mathbf{l}^2-3\left(\mathbf{k}\cdot\mathbf{l}\right)^2}{\left(\mathbf{k}^2+m_{\pi}^2\right)\left(\mathbf{l}^2+m_{\pi}^2\right)}    \right].
\end{align}
Now we define a new coordinate system, $\mathbf{p}_+=(\mathbf{p}_1+\mathbf{p}_2)/2$, $\mathbf{p}_-=(\mathbf{p}_1-\mathbf{p}_2)/2$, $\mathbf{a}=\mathbf{p}_3-\mathbf{p}_+$, $\mathbf{b}=\mathbf{p}_4-\mathbf{p}_+$,
which has Jacobian $\mathop{d^3p}_1\mathop{d^3p}_2\mathop{d^3p}_3\mathop{d^3p}_4=8\mathop{d^3p}_+\mathop{d^3p}_-\mathop{d^3a}\mathop{d^3b}$.  The three-dimensional delta function becomes $\delta^3(\mathbf{a}+\mathbf{b})$, so we integrate over the three-momentum $\mathbf{b}$ and then over the axion energy $\omega$, using the delta function (which became $\delta(\mathbf{p}_-^2-\mathbf{a}^2-m_*\omega)$).  

We are now left with an integral over the three-vectors $\mathbf{p}_+$, $\mathbf{p}_-$, and $\mathbf{a}$, so we choose a coordinate system where $\mathbf{a} = a(0,0,1)$, $\mathbf{p}_- = p_-(\sqrt{1-r^2},0,r)$, and $\mathbf{p}_+ = p_+(\sqrt{1-s^2}\cos{\phi},\sqrt{1-s^2}\sin{\phi},s)$.  Now, $\mathbf{k}^2=p_-^2+a^2-2p_-ar$, $\mathbf{l}^2= p_-^2+a^2+2p_-ar$, $\mathbf{k}\cdot\mathbf{l}=p_-^2-a^2$, $\mathbf{p}_1^2 = p_+^2+p_-^2+2p_+p_-(\sqrt{1-r^2}\sqrt{1-s^2}\cos{\phi}+rs)$, $\mathbf{p}_2^2=p_+^2+p_-^2-2p_+p_-(\sqrt{1-r^2}\sqrt{1-s^2}\cos{\phi}+rs)$, $\mathbf{p}_3^2=p_+^2+a^2+2p_+as$, and $\mathbf{p}_4^2=p_+^2+a^2-2p_+as$.

Now we can integrate over the three trivial angles, giving a factor of $8\pi^2$, leaving us with a six-dimensional integral that we simplify with the coordinate transformations $u = p_+^2/(2m_*T)$, $v=p_-^2/(2m_*T)$, $w=a^2/(2m_*T)$, and we define $\hat{y}=\hat{\mu}/T$.  Thus the key variables in the emissivity expression become
\begin{align}
    \mathbf{k}^2 &= 2m_*T(v+w-2\sqrt{vw}r)\nonumber\\
    \mathbf{l}^2 &= 2m_*T(v+w+2\sqrt{vw}r)\nonumber\\
    \mathbf{k}\cdot\mathbf{l}&=2m_*T(v-w)\nonumber\\
    \beta (E_1-\mu_n) &= -\hat{y}+u+v+2\sqrt{uv}(\sqrt{1-r^2}\sqrt{1-s^2}\cos{\phi}+rs)\\
    \beta (E_2-\mu_n) &= -\hat{y}+u+v-2\sqrt{uv}(\sqrt{1-r^2}\sqrt{1-s^2}\cos{\phi}+rs)\nonumber\\
    \beta (E_3-\mu_n) &= -\hat{y}+u+w+2\sqrt{uw}s\nonumber\\
    \beta (E_4-\mu_n) &= -\hat{y}+u+w-2\sqrt{uw}s.\nonumber
\end{align}
and the axion emissivity is
\begin{align}
    Q &= \frac{32\sqrt{2}}{3\pi^8}\frac{f^4m_n^4G_{an}^2}{m_{\pi}^4}m_*^{1/2}T^{6.5}\int_0^{\infty}\mathop{du}\mathop{dv}\int_0^v\mathop{dw}\int_{-1}^1\mathop{dr}\mathop{ds}\int_0^{2\pi}\mathop{d\phi}u^{1/2}v^{3/2}w^{3/2}(v-w)^2\nonumber\\
    &\times \frac{\left(\alpha^4 \left(r^2+3\right)-6 \alpha^2 \left(r^2-1\right) (v+w)-3 \left(r^2-1\right) \left(2 \left(1-2 r^2\right) v w+v^2+w^2\right)\right)}{\left(2 w \left(\alpha^2-2 r^2
   v+v\right)+\left(\alpha^2+v\right)^2+w^2\right)^2}\nonumber\\
   &\times \left((1+e^{\beta(E_1-\mu_n)})(1+e^{\beta(E_2-\mu_n)})(1+e^{-\beta(E_3-\mu_n)})(1+e^{-\beta(E_4-\mu_n)})\right)^{-1},\label{eq:Q_NR_PS}
\end{align}
where $\alpha = m_{\pi}/\sqrt{2m_*T}$.
This integral can be done numerically.
%%%%%%%%%%%%%%%%%%%%%%%%%%%%%%%%%%%%%%%%%%%%%%%%%
\subsection{Fermi surface approximation and its improvement}
\label{sec:Q_FS_derivation}
In nuclear matter, to good approximation the dominant contribution to a process involving degenerate fermions will be from those fermions near their Fermi surface.  Since calculations of the full phase space integral are often not possible, and almost always result in an integral that must be done numerically, the calculations of the mean free path, rate, and emissivity are almost always done via the Fermi surface approximation, which we describe below.  

In the Fermi surface approximation, the phase space integral (like in Eq.~(\ref{eq:MFP_integral}) and (\ref{eq:emissivity_integral})) is converted into spherical coordinates for each momentum-three vector and then broken up (termed ``phase space decomposition'' \cite{Shapiro:1983du}) into an angular integral $A$ and an integral $J$ over momentum magnitudes (equivalently, energies).  In the angular integral, the fermion momentum magnitudes are set equal to their respective Fermi momenta, while in the energy integral, the energies are integrated over, consistent with thermal blurring of the Fermi surface (although inconsistent with the momenta magnitudes in the angular integral, which were not allowed to vary above or below the Fermi surface).   

The emissivity due to axion emission via $n+n\rightarrow n+n+a$ is given by Eq.~(\ref{eq:emissivity_integral}).  We again neglect the axion 3-momentum in the momentum-conserving delta function and then multiply by one in the form
\begin{align}
    1 &= \int_0^{\infty} \mathop{dp_1}\mathop{dp_2}\mathop{dp_3}\mathop{dp_4} \delta(p_1-p_{Fn})\delta(p_2-p_{Fn})\delta(p_3-p_{Fn})\delta(p_4-p_{Fn})\nonumber\\
    &=\frac{1}{p_{Fn}^4}\int_{m_*+U_n}^{\infty} \mathop{dE_1}\mathop{dE_2}\mathop{dE_3}\mathop{dE_4}E_1^*E_2^*E_3^*E_4^*\delta(p_1-p_{Fn})\delta(p_2-p_{Fn})\delta(p_3-p_{Fn})\delta(p_4-p_{Fn}),
    \label{eq:cleverone}
\end{align}
where we have used $\mathop{dE} = (p/E^*) \mathop{dp}$ as can be seen from the neutron dispersion relation Eq.~(\ref{eq:En}).  
We perform phase space decomposition, obtaining
\begin{equation}
    Q_{FS} = \frac{1}{768\pi^{11}}\frac{G_{an}^2f^4m_n^4}{m_{\pi}^4p_{Fn}^4}A(p_{Fn})J_2(T,\hat{y}).  
\end{equation}
The angular integral $A$ is 
\begin{align}
A(p_{Fn}) &= \int \mathop{d^3p_1}\mathop{d^3p_2}\mathop{d^3p_3}\mathop{d^3p_4} \delta(p_1-p_{Fn})\delta(p_2-p_{Fn})\delta(p_3-p_{Fn})\delta(p_4-p_{Fn})\delta^3(\mathbf{p}_1+\mathbf{p}_2-\mathbf{p}_3-\mathbf{p}_4) \nonumber\\
&\times\left( \frac{\mathbf{k}^4}{\left(\mathbf{k}^2+m_{\pi}^2\right)^2}+\frac{\mathbf{l}^4}{\left(\mathbf{l}^2+m_{\pi}^2\right)^2}+\frac{\mathbf{k}^2\mathbf{l}^2\left(1-3\left(\hat{\mathbf{k}}\cdot \hat{\mathbf{l}}\right)^2\right)}{\left(\mathbf{k}^2+m_{\pi}^2\right)\left(\mathbf{l}^2+m_{\pi}^2\right)}    \right)= 32\pi^3 p_{Fn}^5 F(y),
\label{eq:ang_expression}
\end{align}
where $F(y)$ is given in Eq.~(\ref{eq:Fofy}).  The energy integral is 
\begin{equation}
    J_2(T) = \int\mathop{d^3\omega} \int_{m_*+U_n}^{\infty} \mathop{dE_1}\mathop{dE_2}\mathop{dE_3}\mathop{dE_4}\delta(E_1+E_2-E_3-E_4-\omega)f_1f_2(1-f_3)(1-f_4).
    \label{eq:J2_original}
\end{equation}
The energy integral is evaluated by changing to dimensionless variables centered at the Fermi energy $x_i = (E_i - \mu_n)/T$, and then to variables $u=x_1+x_2$ and $v=x_1-x_2$ and so Eq.~(\ref{eq:J2_original}) becomes
\begin{align}
   J_2(T,\hat{y}) &\equiv T^3\int\mathop{d^3\omega}\int_{-\hat{y}}^{\infty}\mathop{dx_1}\mathop{dx_2}\mathop{dx_3}\mathop{dx_4}\frac{\delta(x_1+x_2-x_3-x_4-\omega/T)}{(1+e^{x_1})(1+e^{x_2})(1+e^{-x_3})(1+e^{-x_4})}\nonumber\\
    &= 8\pi T^6 \int_{-\hat{y}}^{\infty}\mathop{dx_1}\mathop{dx_2}\int_0^{x_1+x_2+2\hat{y}}\mathop{dz}z^2\frac{\ln{\left(\frac{\cosh{((x_1+x_2-z+\hat{y})/2)}}{\cosh{(\hat{y}/2)}}\right)}}{(1+e^{x_1})(1+e^{x_2})(1-e^{z-x_1-x_2})}
    \label{eq:J2}\nonumber\\
    &= 16\pi T^6\int_{-2\hat{y}}^{\infty}\mathop{du}\frac{1}{1-e^u}\ln{\left\{\frac{\cosh{(\hat{y}/2)}}{\cosh{\left[(u+\hat{y})/2\right]}}\right\}}\int_0^{u+2\hat{y}}\mathop{dw}\frac{w^2}{1-e^{w-u}}\ln{\left\{\frac{\cosh{\left[(u+\hat{y}-w)/2\right]}}{\cosh{(\hat{y}/2)}}\right\}}\\
    &= 16\pi T^6K_2(\hat{y})\nonumber
\end{align}
where $\hat{y} = (\mu_n^*-m_*)/T$ and $K_2(\hat{y})$ is given in Eq.~(\ref{K2}).  For strongly degenerate nuclear matter ($\hat{y}\rightarrow\infty$), 
\begin{equation}
    K_2(\hat{y}\rightarrow\infty) = \frac{31}{1890}\pi^6.
\end{equation}

%%%%%%%%%%%%%%%%%%%%%%%%%%%%%%%%%%
\subsection{Comparison of emissivity expressions}
\begin{figure*}[t!]
\begin{minipage}[t]{0.5\linewidth}
\includegraphics[width=.95\linewidth]{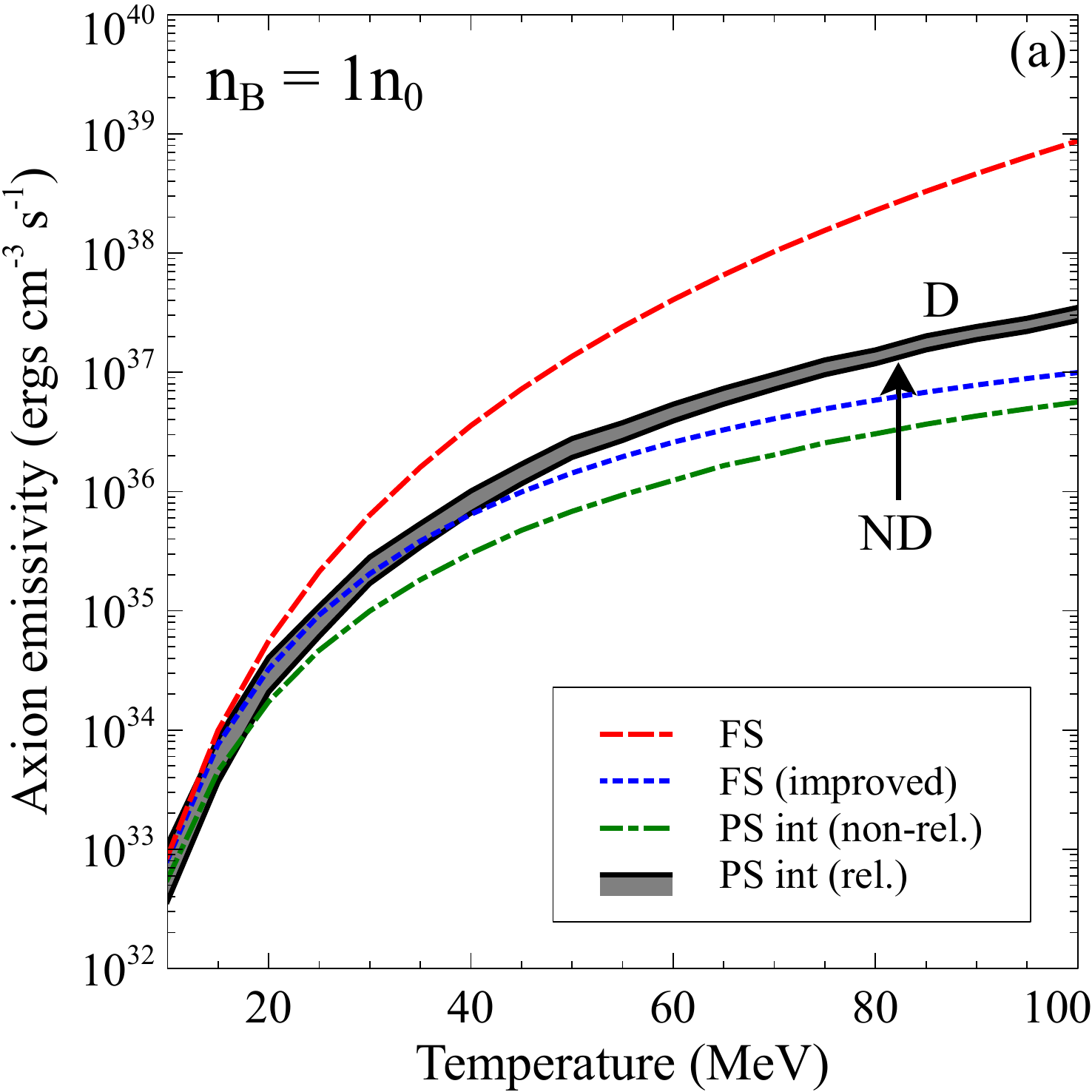}
\end{minipage}\hfill%
\begin{minipage}[t]{0.5\linewidth}
\includegraphics[width=.95\linewidth]{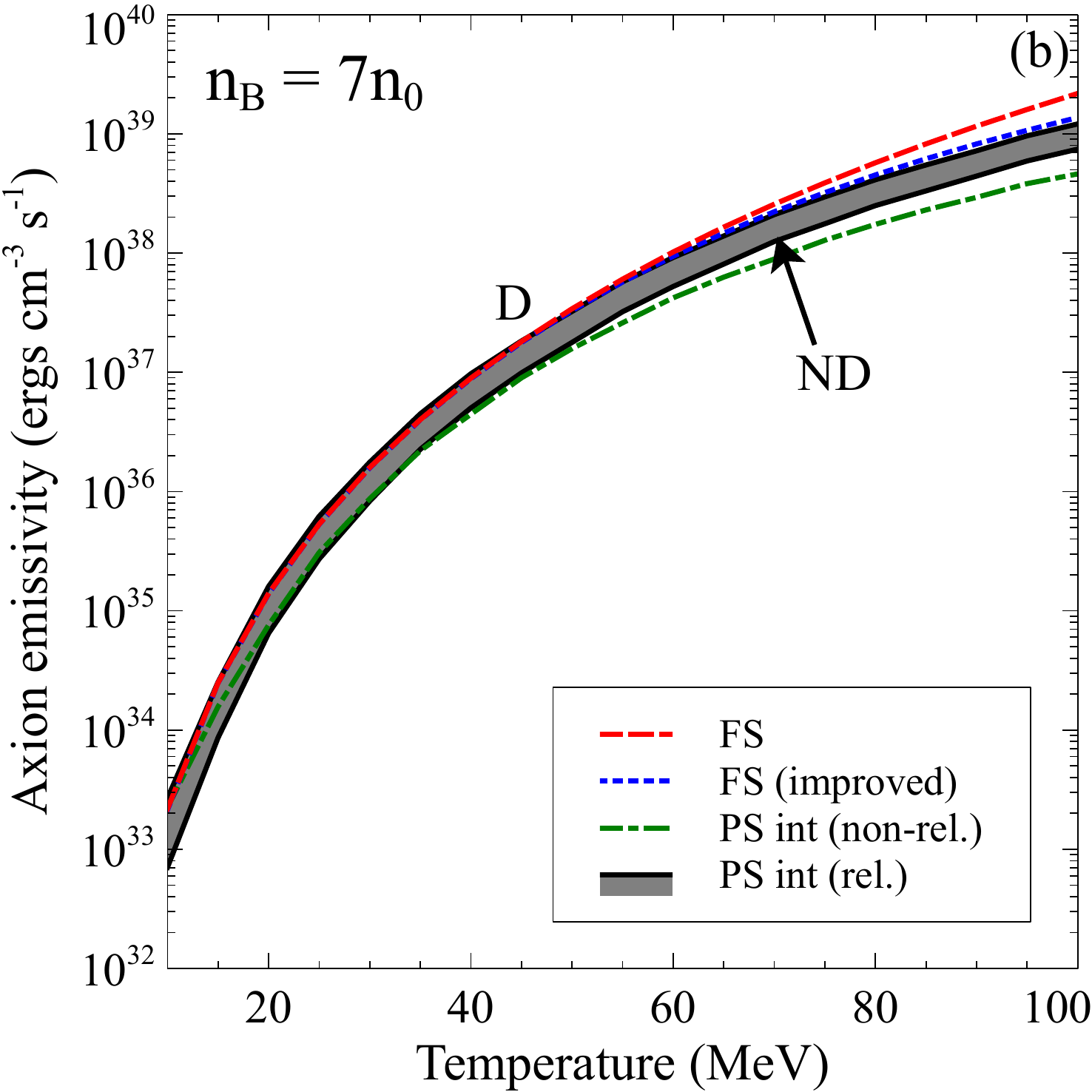}
\end{minipage}%
\caption{Comparison of several approximations of the axion emissivity, including the FS approximation [Eq.~(\ref{eq:Q_FS})] (red, dashed), our improvement to the FS approximation [Eq.~(\ref{eq:Q_new})] (blue, dotted), the non-relativistic phase space integral [Eq.~(\ref{eq:Q_NR_PS})] (green, dash-dotted), and the fully relativistic phase space integral (with constant matrix element) [Eq.~(\ref{eq:exact_emissivity})] (black, solid) at densities of $1n_0$ (a) and $7n_0$ (b).  The axion-neutron coupling is chosen to be $G_{an} = G_{\text{SN1987A}}.$}
\label{fig:compare_Q}
\end{figure*}

In Fig.~\ref{fig:compare_Q} we compare various approximations of the axion emissivity, namely, the FS approximation [Eq.~(\ref{eq:Q_FS})], our improvement to the FS approximation [Eq.~(\ref{eq:Q_new})], the non-relativistic phase space integral [Eq.~(\ref{eq:Q_NR_PS})], and the fully relativistic phase space integral\footnote{In the fully relativistic phase space integral, we choose two values of $k_{\text{typ}}$: $k_{\text{typ}}^2=3m_*T$ and also $k_{\text{typ}}^2=p_{Fn}^2$ as upper and lower bounds, marked ``ND'' and ``D'' respectively in Fig.~\ref{fig:compare_Q}.} (with constant matrix element) [Eq.~(\ref{eq:exact_emissivity})].  We see that at $1n_0$, at low temperature the approximations all agree.  Here the neutrons are degenerate, which explains the success of the FS approximation, and the neutrons are indeed nonrelativistic because their Fermi momentum is not yet large, which explains the success of the nonrelativistic approximation.  As the temperature increases, the neutrons become non-degenerate and so the FS approximation fails dramatically.  The improved treatment of the FS approximation does better than the original, but still fails at temperatures above 40-50 MeV.  The NR approximation fails at high temperature because it neglects the axion momentum in the 3d delta function (at $T=100 \text{ MeV}$, the axion spectrum peaks at $\omega = 300 \text{ MeV}$, which is not negligible compared to the neutron Fermi momentum of 320 MeV).  At $7n_0$, the neutrons are always degenerate, and thus the Fermi surface approximation and its improvement match the full phase space integral quite well.  The non-relativistic phase space integral doesn't match as well because the approximation $E^* \approx m_*$ in the denominator of the phase space integral becomes a very poor approximation as the effective mass dwindles to around 250 MeV at this high density.

%%%%%%%%%%%%%%%%%%%%%%%%%%%%%%%%%%%%%%%%%%%%%%%%%%
\section{Axion mean free path integrals}
Below, we detail a series of approximations for the axion mean free path, Eq.~(\ref{eq:MFP_integral}).
\subsection{Relativistic, constant-matrix-element phase space integration}
\label{sec:rel_PS_MFP}
Following the same procedure as for the emissivity full phase space integral, we can do the full phase space integration of the axion mean free path (with the same approximation of a constant matrix element) and we obtain
\begin{align}
    &\lambda^{-1} = \left(1-\frac{\beta}{3}\right) \frac{f^4 m_n^4 G_{an}^2}{4\pi^5 m_{\pi}^4\omega} \left(1+\frac{m_{\pi}^2}{k_{\text{typ}}^2}\right)^{-2}\int_{m_*}^{\infty}\mathop{dq_0}\int_0^{\infty}\mathop{dq}\int_{-1}^1\mathop{dr}\int_0^{\sqrt{(q_0+\omega)^2-m_*^2}}\mathop{dx}\int_0^{\sqrt{q_0^2-m_*^2}}\mathop{dk}\nonumber\\
    &\times \frac{kqx(q_0-\sqrt{k^2+m_*^2})\theta(2kq-\vert q_0^2-q^2-2q_0\sqrt{k^2+m_*^2} \vert)}{\sqrt{x^2+m_*^2}\sqrt{k^2+m_*^2}\sqrt{\omega^2+q^2+2\omega qr}\sqrt{k^2+m_*^2+q_0^2-2q_0\sqrt{k^2+m_*^2}}(1+e^{(\sqrt{k^2+m_*^2}-\mu_n^*)/T})}\label{eq:MFPexactanswer}\\
    &\times \frac{\theta(2x\sqrt{\omega^2+q^2+2\omega qr}-\vert q^2+2\omega qr-q_0^2-2q_0\omega +2(q_0+\omega)\sqrt{x^2+m_*^2} \vert)}{(1+e^{(\sqrt{k^2+m_*^2+q_0^2-2q_0\sqrt{k^2+m_*^2}}-\mu_n^*)/T})(1+e^{-(\sqrt{x^2+m_*^2}-\mu_n^*)/T})(1+e^{-(\omega + q_0 - \sqrt{x^2+m_*^2}-\mu_n^*)/T})}.\nonumber
\end{align}
%%%%%%%%%%%%%%%%%%%%%%%%%%%%%%%%%%%%%%%%%%%%%%%%%%%%%%%%%%%
\subsection{Non-relativistic phase space integration}
\label{sec:MFP_nonrel_PS}
The axion mean free path [Eq.~(\ref{eq:MFP_integral})] can be computed assuming non-relativistic neutrons.  A similar calculation (for nondegenerate neutrons) has been done in \cite{PhysRevD.42.3297,Giannotti:2005tn} and was extended to arbitrary degeneracy in \cite{Carenza:2019pxu}.  We start with Eq.~(\ref{eq:MFP_integral}), keeping the matrix element as momentum-dependent and thus inside the integral.  We use the nonrelativistic (quadratic) approximation for the neutron dispersion relations except in the four energy denominators where $E^* = m_*$, just as in the emissivity calculation.  We obtain
\begin{align}
    \lambda_a^{-1} &= \frac{1}{48\pi^{8}}\frac{f^4 m_n^4 G_{an}^2}{m_{\pi}^4\omega m_*^3}\int \mathop{d^3p_1}\mathop{d^3p_2}\mathop{d^3p_3}\mathop{d^3p_4}\delta(p_1^2+p_2^2-p_3^2-p_4^2+2m_*\omega)\delta^3(\mathbf{p}_1+\mathbf{p}_2-\mathbf{p}_3-\mathbf{p}_4)\nonumber\\
    &\times f_1f_2(1-f_3)(1-f_4)\left[ \frac{\mathbf{k}^4}{\left(\mathbf{k}^2+m_{\pi}^2\right)^2}+\frac{\mathbf{l}^4}{\left(\mathbf{l}^2+m_{\pi}^2\right)^2}+\frac{\mathbf{k}^2\mathbf{l}^2-3\left(\mathbf{k}\cdot\mathbf{l}\right)^2}{\left(\mathbf{k}^2+m_{\pi}^2\right)\left(\mathbf{l}^2+m_{\pi}^2\right)}    \right].
\end{align}
Just as in the emissivity calculation, we transform coordinates to $\{\mathbf{p}_+,\mathbf{p}_-,\mathbf{a},\mathbf{b}\}$, picking up a factor of 8 from the Jacobian, and then we integrate over $\mathbf{b}$, using the three-dimensional delta function $\delta^3(\mathbf{a}+\mathbf{b})$.  We choose the same coordinate system as in the emissivity calculation (Appendix \ref{sec:Q_nonrel_PS}), aligning the ``z'' axis along the $\mathbf{a}$ three-momentum vector.  We integrate over the 3 trivial angles and then use the energy delta function to integrate over $a$.  Then we create nondimensional variables $u=p_+^2/(2m_*T)$ and $v=p_-^2/(2m_*T)$, and define $\hat{y}=\hat{\mu}/T=(\mu^*-m_*)/T$ as in the emissivity calculation.  We also define $\gamma = \omega/(2T)$.
\begin{align}
    \mathbf{k}^2 &= 2m_*T(2v+\gamma-2\sqrt{v}\sqrt{v+\gamma}r)\nonumber\\
    \mathbf{l}^2 &= 2m_*T(2v+\gamma+2\sqrt{v}\sqrt{v+\gamma}r)\nonumber\\
    \mathbf{k}\cdot\mathbf{l} &= -m_*\omega\nonumber\\
    \beta(E_1-\mu_n) &= -\hat{y}+u+v+2\sqrt{uv}(\sqrt{1-r^2}\sqrt{1-s^2}\cos{\phi}+rs)\\
    \beta(E_2-\mu_n) &= -\hat{y}+u+v-2\sqrt{uv}(\sqrt{1-r^2}\sqrt{1-s^2}\cos{\phi}+rs)\nonumber\\
    \beta(E_3-\mu_n) &= -\hat{y}+u+v+\gamma + 2\sqrt{u}\sqrt{v+\gamma}s\nonumber\\
    \beta(E_4-\mu_n) &= -\hat{y}+u+v+\gamma-2\sqrt{u}\sqrt{v+\gamma}s\nonumber.
\end{align}
We find that
\begin{align}
    \lambda^{-1} &= \frac{8\sqrt{2}}{3\pi^6}\frac{f^4m_n^4G_{an}^2}{m_{\pi}^4\omega}m_*^{1/2}T^{3.5}\int_0^{\infty}\mathop{du}\mathop{dv}\int_{-1}^1\mathop{dr}\mathop{ds}\int_0^{2\pi}\mathop{d\phi}u^{1/2}v^{3/2}(v+\gamma)^{3/2}\label{eq:MFP_NR_PS}\\
    &\times \frac{\alpha ^4 \left(r^2+3\right)-6 \alpha ^2 \left(r^2-1\right) (\gamma +2 v)+3 \left(r^2-1\right) \left(-\gamma ^2+4 \gamma  \left(r^2-1\right) v+4 \left(r^2-1\right) v^2\right)}{\left(4 v \left(\alpha
   ^2-\gamma  r^2+\gamma \right)+\left(\alpha ^2+\gamma \right)^2-4 \left(r^2-1\right) v^2\right)^2}.\nonumber\\
   &\times\left((1+e^{\beta(E_1-\mu_n)})(1+e^{\beta(E_2-\mu_n)})(1+e^{-\beta(E_3-\mu_n)})(1+e^{-\beta(E_4-\mu_n)})\right)^{-1}.\nonumber
\end{align}
%%%%%%%%%%%%%%%%%%%%%%%%%%%%%%%%%%%%%%%%%%%%%%%%%%
\subsection{Fermi surface approximation and its improvement}
\label{sec:mfp_FS_calculation}
The mean free path of an axion due to the process $n+n+a\rightarrow n+n$ is given by Eq.~(\ref{eq:MFP_integral}).  We neglect the 3-momentum of the axion in the momentum-conserving delta function, and then multiply by one in the form Eq.~(\ref{eq:cleverone}).  Then we perform phase space decomposition, splitting the integral into integral expressions $A$ and $J_1$
\begin{equation}
    \lambda^{-1} = \frac{f^4G_{an}^2m_n^4}{96\pi^8\omega m_{\pi}^4 p_{Fn}^4}A(p_{Fn})J_1(\omega,T,\hat{y}),
\end{equation}
The angular integral is the same as for the emissivity calculation (see Eq.~(\ref{eq:ang_expression})) and the energy integral is
\begin{equation}
    J_1(\omega,T,\hat{y}) = \int_{m_*+U_n}^{\infty} \mathop{dE_1}\mathop{dE_2}\mathop{dE_3}\mathop{dE_4}\delta(E_1+E_2-E_3-E_4+\omega)f_1f_2(1-f_3)(1-f_4).
    \label{eq:J1_original}
\end{equation}
The energy integral is evaluated by changing to dimensionless variables centered at the Fermi energy $x_i = (E_i - \mu_n)/T$, and then to variables $u=x_1+x_2$ and $v=x_1-x_2$ and so Eq.~(\ref{eq:J1_original}) becomes
\begin{align}
    J_1(\omega,T,\hat{y}) &\equiv T^3 \int_{-\hat{y}}^{\infty}\mathop{dx_1}\mathop{dx_2}\mathop{dx_3}\mathop{dx_4}\frac{\delta(x_1+x_2-x_3-x_4+\omega/T)}{(1+e^{x_1})(1+e^{x_2})(1+e^{-x_3})(1+e^{-x_4})}\nonumber\\
    &= 2T^3\int_{-\hat{y}}^{\infty}\mathop{dx_1}{dx_2}\frac{\ln{\left(\frac{\cosh{((x_1+x_2+\omega/T+\hat{y})/2})}{\cosh{(\hat{y}/2)}}\right)}}{(1+e^{x_1})(1+e^{x_2})(1-e^{-x_1-x_2-\omega/T})}\nonumber\\
    &= 4T^3 K_1(\hat{y},\omega/T),
\end{align}
where $\hat{y} = (\mu_n^*-m_*)/T$ and $K_1(\hat{y},\omega/T)$ is given in Eq.~(\ref{eq:K1}).
In the literature, it is standard to consider strongly degenerate nuclear matter where $\hat{y}\rightarrow\infty$ and so
\begin{equation}
    K_1(\hat{y}\rightarrow\infty,\omega/T) = \left(\frac{\omega/T}{24}\right)\frac{(\omega/T)^2+4\pi^2}{1-e^{-\omega/T}},
\end{equation}
and we arrive at the formula for the axion mean free path in strongly degenerate nuclear matter, Eq.~(\ref{eq:lambdaFS}).  In semi-degenerate matter, the improved FS approximation yields Eq.~(\ref{eq:lambda_new}) for the axion mean free path.
%%%%%%%%%%%%%%%%%%%%%%%%%%%%%%%%%%%%%%%%%%%%%%%%%%%%%%%%%%%%%
\subsection{Comparison of MFP expressions}
\begin{figure*}[t!]
\begin{minipage}[t]{0.5\linewidth}
\includegraphics[width=.95\linewidth]{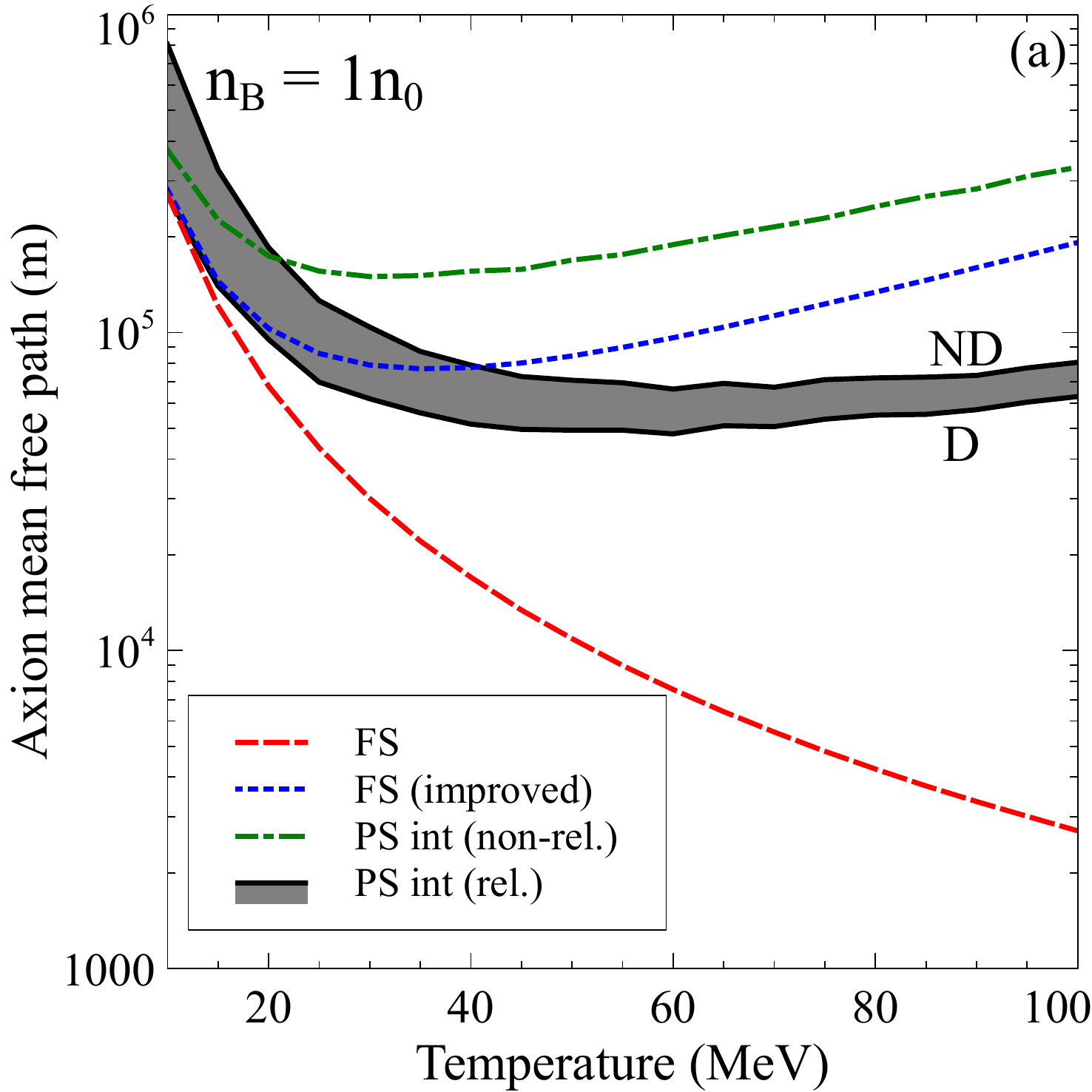}
\end{minipage}\hfill%
\begin{minipage}[t]{0.5\linewidth}
\includegraphics[width=.95\linewidth]{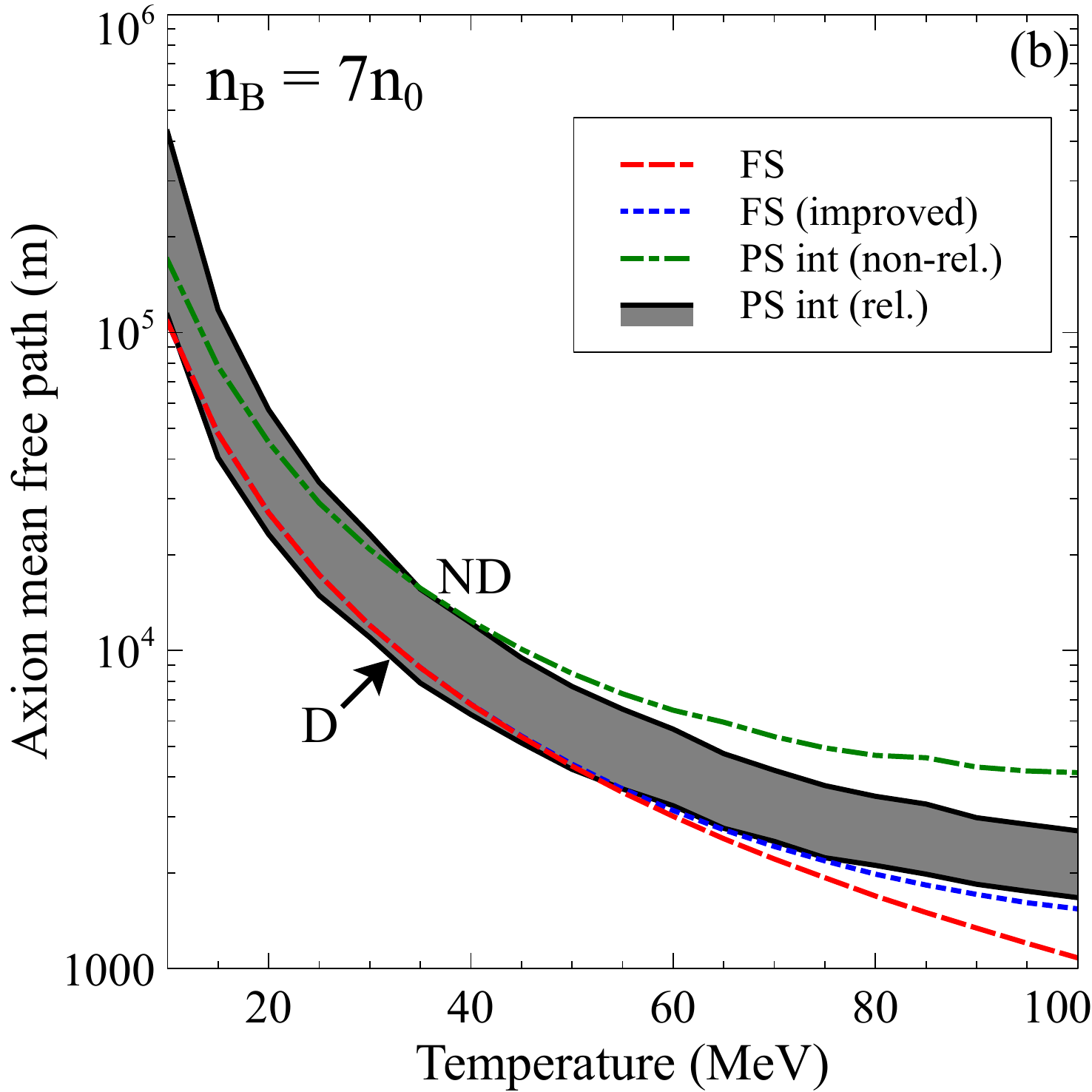}
\end{minipage}%
\caption{Comparison of the various approximations of the axion MFP including the FS approximation [Eq.~(\ref{eq:lambdaFS})] (red, dashed), our improvement to the FS approximation [Eq.~(\ref{eq:lambda_new})] (blue, dotted), the non-relativistic phase space integral [Eq.~(\ref{eq:MFP_NR_PS})] (green, dash-dotted), and the fully relativistic phase space integral (with constant matrix element) [Eq.~(\ref{eq:MFPexactanswer})] (black, solid) at densities of $1n_0$ (a) and $7n_0$ (b).  The axion-neutron coupling is chosen to be $G_{an} = G_{\text{SN1987A}}.$}
\label{fig:MFP_comparison}
\end{figure*}
In Fig.~\ref{fig:MFP_comparison} we compare different approximations for the axion mean free path, namely, the FS approximation [Eq.~(\ref{eq:lambdaFS})], our improvement to the FS approximation [Eq.~(\ref{eq:lambda_new})], the non-relativistic phase space integral [Eq.~(\ref{eq:MFP_NR_PS})], and the fully relativistic phase space integral\footnote{In the fully relativistic phase space integral, we choose two values of $k_{\text{typ}}$: $k_{\text{typ}}^2=3m_*T$ and also $k_{\text{typ}}^2=p_{Fn}^2$ as upper and lower bounds, marked ``ND'' and ``D'' respectively in Fig.~\ref{fig:MFP_comparison}.} (with constant matrix element) [Eq.~(\ref{eq:MFPexactanswer})]. 

 We see in Fig.~\ref{fig:MFP_comparison} that at $1n_0$, at low temperature the approximations all agree (though the nonrelativistic phase space integral deviates slightly from the rest, because of the approximation in the energy denominators $E^*\approx m_*$, which shows up more in this figure than in Fig.~\ref{fig:compare_Q} because of the difference in y-axis scales).  At saturation density and low temperature, neutrons are degenerate, which explains the success of the FS approximation, and the neutrons are indeed nonrelativistic because their Fermi momentum is not yet large, which explains the success of the nonrelativistic approximation.  As the temperature increases, the neutrons become non-degenerate and so the FS approximation fails dramatically.  The improved treatment of the FS approximation does better than the original, but still fails at temperatures above about 40 MeV.  The NR approximation fails at high temperature because it neglects the axion momentum in the 3d delta function (again, at $T=100 \text{ MeV}$, the axion spectrum peaks at $\omega = 300 \text{ MeV}$, which is not negligible compared the neutron Fermi momentum of 320 MeV).  At $7n_0$, the neutrons are always degenerate, and thus the Fermi surface approximation and its improvement match the full phase space integral quite well.  Again, the non-relativistic phase space integral doesn't match as well because the approximation $E^* \approx m_*$ in the denominator of the phase space integral becomes a very poor approximation as the effective mass dwindles to around 250 MeV.
 %%%%%%%%%%%%%%%%%%%%%%%%%%%%%%%%%%%%%%
 \section{Curve fits}
 \label{sec:curve_fits}
 \begin{figure}[tbp]
 \centering
\includegraphics[width=.45\textwidth]{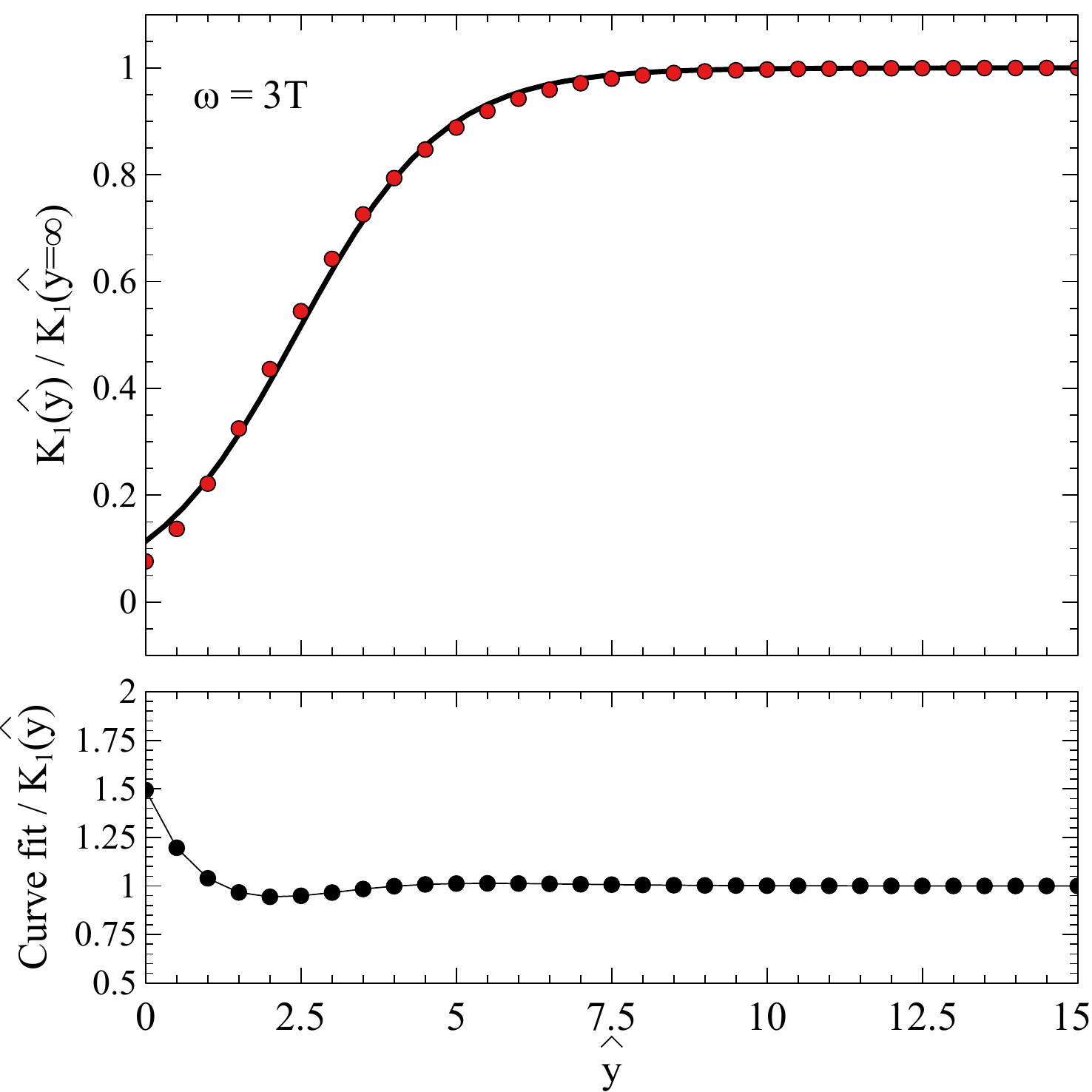}
\caption{Curve fit plotted over the integral $K_1(\hat{y},\omega/T=3)$.  The bottom panels show the error of the curve fit.  The curve fit is only valid for $\hat{y}\gtrsim0$.}
\label{fig:K1_vs_b}
\end{figure}
\begin{figure}[tbp]
\centering
\includegraphics[width=.45\textwidth]{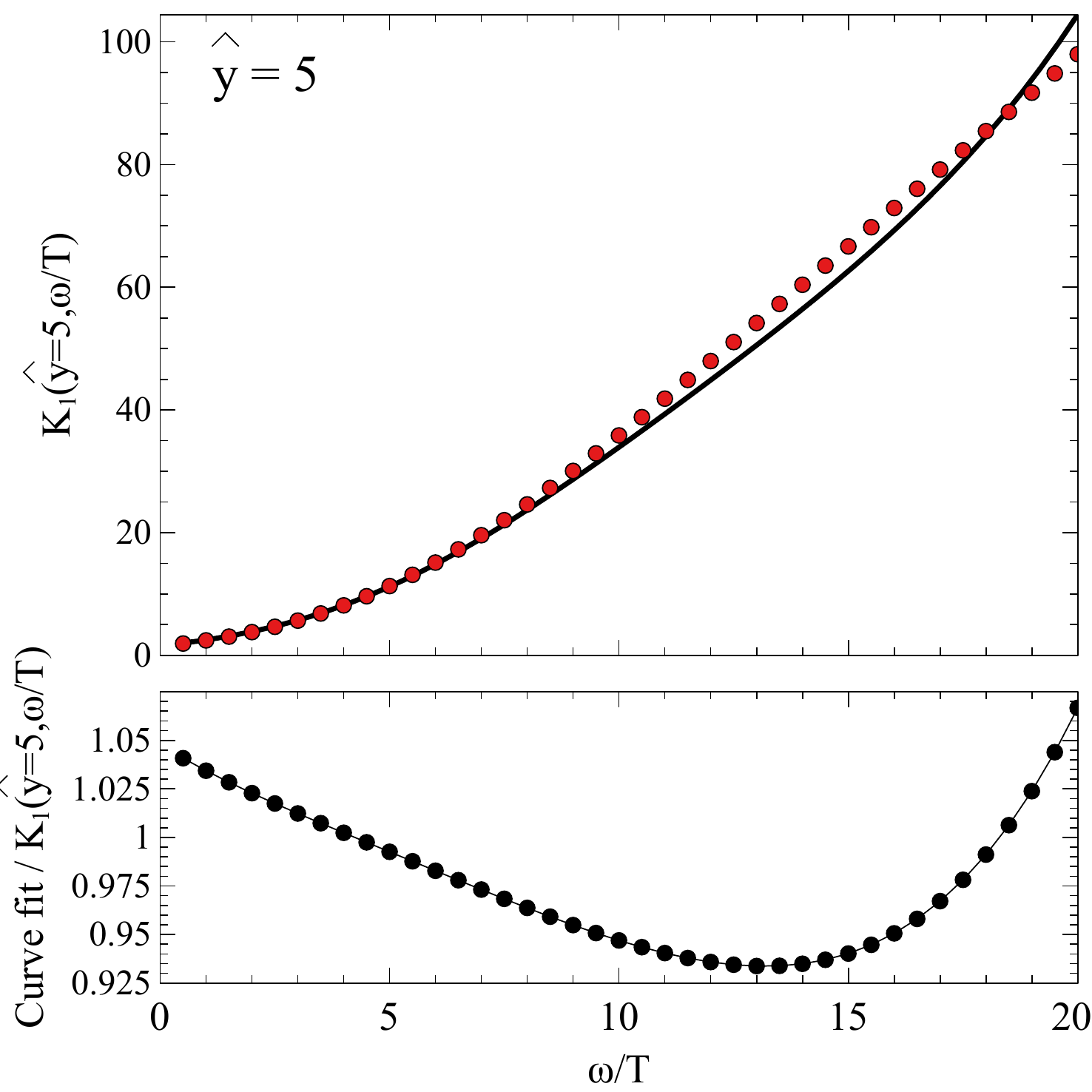}
\caption{Plot of the curve fit $K_1(\hat{y}=5,\omega/T)$ on top of the exact evaluation.  The error is plotted in the lower panel.  This fit is valid for $0<\omega/T\lesssim 20$.}
\label{fig:K1_vs_w}
\end{figure}
\begin{figure}
\includegraphics[width=.45\textwidth]{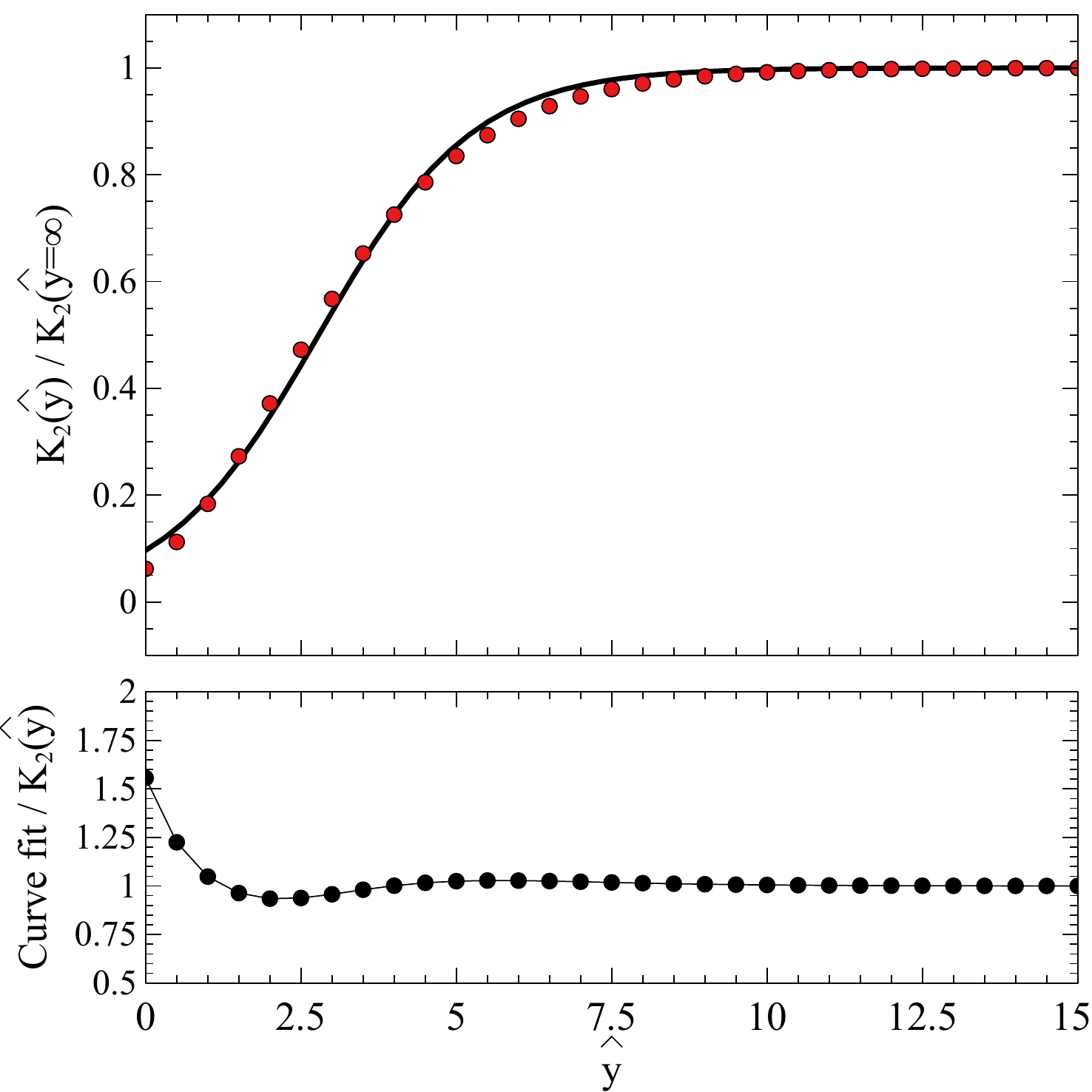}[tbp]
\centering
\caption{Curve fit plotted over the integral $K_2(\hat{y})$.  The bottom panels show the error of the curve fit.  The curve fit is only valid for $\hat{y}\gtrsim0$.}
\label{fig:K2_vs_b}
\end{figure}
The integrals $K_1(\hat{y},\omega/T)$ [Eq.~(\ref{eq:K1})] and $K_2(\hat{y})$ [Eq.~(\ref{K2})] can be fit to functions proportional to Fermi-Dirac factors with arguments $\hat{y}$.  We find
\begin{align}
     K_1(\hat{y},\omega/T) &= \frac{K_1(\hat{y}=\infty,\omega/T)}{1+e^{-(\alpha-\beta\omega/T)(\hat{y}-\gamma\omega/T-\delta)}}\label{eq:K1_curve_fit}\\
     \alpha &= 0.9322 \pm 0.0012\nonumber\\
     \beta &= 0.0277 \pm 0.0001\nonumber\\
     \gamma &= 0.2945 \pm 0.0002\nonumber\\
     \delta &= 1.5372 \pm 0.0024\nonumber
\end{align}
and
\begin{align}
    K_2(T,\hat{y}) &= \frac{K_2(T,\hat{y}=\infty)}{1+e^{-\epsilon(\hat{y}-\zeta)}}\label{eq:K2_curve_fit}\\
    \epsilon &= 0.8032\pm 0.0087\nonumber\\
    \zeta &= 2.7844\pm 0.0154\nonumber
\end{align}
Comparisons of the curve fits to the original numerical integrals are found in Figs.~\ref{fig:K1_vs_b}, \ref{fig:K1_vs_w}, and \ref{fig:K2_vs_b}.
%%%%%%%%%%%%%%%%%%%%%%%%%%%%%%%%%
\bibliographystyle{JHEP}
\bibliography{axion}{}

\providecommand{\href}[2]{#2}\begingroup\raggedright\begin{thebibliography}{10}

\bibitem{Chang:2018rso}
J.~H. Chang, R.~Essig and S.~D. McDermott, \emph{{Supernova 1987A Constraints
  on Sub-GeV Dark Sectors, Millicharged Particles, the QCD Axion, and an
  Axion-like Particle}},
  \href{http://dx.doi.org/10.1007/JHEP09(2018)051}{\emph{JHEP} {\bf 09} (2018)
  051}, [\href{https://arxiv.org/abs/1803.00993}{{\tt 1803.00993}}].

\bibitem{Raffelt:1996wa}
G.~G. Raffelt, \emph{{Stars as laboratories for fundamental physics}}.
\newblock 1996.

\bibitem{TheLIGOScientific:2017qsa}
{\scshape LIGO Scientific, Virgo} collaboration, B.~P. Abbott et~al.,
  \emph{{GW170817: Observation of Gravitational Waves from a Binary Neutron
  Star Inspiral}},
  \href{http://dx.doi.org/10.1103/PhysRevLett.119.161101}{\emph{Phys. Rev.
  Lett.} {\bf 119} (2017) 161101},
  [\href{https://arxiv.org/abs/1710.05832}{{\tt 1710.05832}}].

\bibitem{Lucca:2019ohp}
M.~Lucca and L.~Sagunski, \emph{{The lifetime of binary neutron star merger
  remnants}},  \href{https://arxiv.org/abs/1909.08631}{{\tt 1909.08631}}.

\bibitem{Baiotti:2016qnr}
L.~Baiotti and L.~Rezzolla, \emph{{Binary neutron star mergers: a review of
  Einstein’s richest laboratory}},
  \href{http://dx.doi.org/10.1088/1361-6633/aa67bb}{\emph{Rept. Prog. Phys.}
  {\bf 80} (2017) 096901}, [\href{https://arxiv.org/abs/1607.03540}{{\tt
  1607.03540}}].

\bibitem{Gill:2019bvq}
R.~Gill, A.~Nathanail and L.~Rezzolla, \emph{{When Did the Remnant of GW170817
  Collapse to a Black Hole?}},
  \href{http://dx.doi.org/10.3847/1538-4357/ab16da}{\emph{Astrophys. J.} {\bf
  876} (2019) 139}, [\href{https://arxiv.org/abs/1901.04138}{{\tt
  1901.04138}}].

\bibitem{Hanauske:2019qgs}
M.~Hanauske, J.~Steinheimer, A.~Motornenko, V.~Vovchenko, L.~Bovard, E.~R. Most
  et~al., \emph{{Neutron Star Mergers: Probing the EoS of Hot, Dense Matter by
  Gravitational Waves}},
  \href{http://dx.doi.org/10.3390/particles2010004}{\emph{Particles} {\bf 2}
  (2019) 44--56}.

\bibitem{Radice:2018pdn}
D.~Radice, A.~Perego, K.~Hotokezaka, S.~A. Fromm, S.~Bernuzzi and L.~F.
  Roberts, \emph{{Binary Neutron Star Mergers: Mass Ejection, Electromagnetic
  Counterparts and Nucleosynthesis}},
  \href{http://dx.doi.org/10.3847/1538-4357/aaf054}{\emph{Astrophys. J.} {\bf
  869} (2018) 130}, [\href{https://arxiv.org/abs/1809.11161}{{\tt
  1809.11161}}].

\bibitem{Radice:2016rys}
D.~Radice, S.~Bernuzzi, W.~Del~Pozzo, L.~F. Roberts and C.~D. Ott,
  \emph{{Probing Extreme-Density Matter with Gravitational Wave Observations of
  Binary Neutron Star Merger Remnants}},
  \href{http://dx.doi.org/10.3847/2041-8213/aa775f}{\emph{Astrophys. J.} {\bf
  842} (2017) L10}, [\href{https://arxiv.org/abs/1612.06429}{{\tt
  1612.06429}}].

\bibitem{Wanajo:2014wha}
S.~Wanajo, Y.~Sekiguchi, N.~Nishimura, K.~Kiuchi, K.~Kyutoku and M.~Shibata,
  \emph{{Production of all the $r$-process nuclides in the dynamical ejecta of
  neutron star mergers}},
  \href{http://dx.doi.org/10.1088/2041-8205/789/2/L39}{\emph{Astrophys. J.}
  {\bf 789} (2014) L39}, [\href{https://arxiv.org/abs/1402.7317}{{\tt
  1402.7317}}].

\bibitem{Kasen:2013xka}
D.~Kasen, N.~R. Badnell and J.~Barnes, \emph{{Opacities and Spectra of the
  $r$-process Ejecta from Neutron Star Mergers}},
  \href{http://dx.doi.org/10.1088/0004-637X/774/1/25}{\emph{Astrophys. J.} {\bf
  774} (2013) 25}, [\href{https://arxiv.org/abs/1303.5788}{{\tt 1303.5788}}].

\bibitem{Metzger:2017wot}
B.~D. Metzger, \emph{{Welcome to the Multi-Messenger Era! Lessons from a
  Neutron Star Merger and the Landscape Ahead}},
  \href{https://arxiv.org/abs/1710.05931}{{\tt 1710.05931}}.

\bibitem{Shibata:2006nm}
M.~Shibata and K.~Taniguchi, \emph{{Merger of binary neutron stars to a black
  hole: disk mass, short gamma-ray bursts, and quasinormal mode ringing}},
  \href{http://dx.doi.org/10.1103/PhysRevD.73.064027}{\emph{Phys. Rev.} {\bf
  D73} (2006) 064027}, [\href{https://arxiv.org/abs/astro-ph/0603145}{{\tt
  astro-ph/0603145}}].

\bibitem{Baiotti:2008ra}
L.~Baiotti, B.~Giacomazzo and L.~Rezzolla, \emph{{Accurate evolutions of
  inspiralling neutron-star binaries: prompt and delayed collapse to black
  hole}}, \href{http://dx.doi.org/10.1103/PhysRevD.78.084033}{\emph{Phys. Rev.}
  {\bf D78} (2008) 084033}, [\href{https://arxiv.org/abs/0804.0594}{{\tt
  0804.0594}}].

\bibitem{Hotokezaka:2011dh}
K.~Hotokezaka, K.~Kyutoku, H.~Okawa, M.~Shibata and K.~Kiuchi, \emph{{Binary
  Neutron Star Mergers: Dependence on the Nuclear Equation of State}},
  \href{http://dx.doi.org/10.1103/PhysRevD.83.124008}{\emph{Phys. Rev.} {\bf
  D83} (2011) 124008}, [\href{https://arxiv.org/abs/1105.4370}{{\tt
  1105.4370}}].

\bibitem{Dietrich:2019shr}
T.~Dietrich and K.~Clough, \emph{Cooling binary neutron star remnants via
  nucleon-nucleon-axion bremsstrahlung},
  \href{http://dx.doi.org/10.1103/PhysRevD.100.083005}{\emph{Phys.Rev.D} {\bf
  100} (2019) 083005}, [\href{https://arxiv.org/abs/1909.01278}{{\tt
  1909.01278}}].

\bibitem{Brinkmann:1988vi}
R.~P. Brinkmann and M.~S. Turner, \emph{{Numerical Rates for Nucleon-Nucleon
  Axion Bremsstrahlung}},
  \href{http://dx.doi.org/10.1103/PhysRevD.38.2338}{\emph{Phys. Rev.} {\bf D38}
  (1988) 2338}.

\bibitem{Radice:2016dwd}
D.~Radice, F.~Galeazzi, J.~Lippuner, L.~F. Roberts, C.~D. Ott and L.~Rezzolla,
  \emph{{Dynamical Mass Ejection from Binary Neutron Star Mergers}},
  \href{http://dx.doi.org/10.1093/mnras/stw1227}{\emph{Mon. Not. Roy. Astron.
  Soc.} {\bf 460} (2016) 3255--3271},
  [\href{https://arxiv.org/abs/1601.02426}{{\tt 1601.02426}}].

\bibitem{Graham:2015ouw}
P.~W. Graham, I.~G. Irastorza, S.~K. Lamoreaux, A.~Lindner and K.~A. van
  Bibber, \emph{{Experimental Searches for the Axion and Axion-Like
  Particles}},
  \href{http://dx.doi.org/10.1146/annurev-nucl-102014-022120}{\emph{Ann. Rev.
  Nucl. Part. Sci.} {\bf 65} (2015) 485--514},
  [\href{https://arxiv.org/abs/1602.00039}{{\tt 1602.00039}}].

\bibitem{PhysRevD.42.3297}
A.~Burrows, M.~T. Ressell and M.~S. Turner, \emph{Axions and sn 1987a: Axion
  trapping}, \href{http://dx.doi.org/10.1103/PhysRevD.42.3297}{\emph{Phys. Rev.
  D} {\bf 42} (Nov, 1990) 3297--3309}.

\bibitem{PhysRevD.39.1020}
A.~Burrows, M.~S. Turner and R.~P. Brinkmann, \emph{Axions and sn 1987a},
  \href{http://dx.doi.org/10.1103/PhysRevD.39.1020}{\emph{Phys. Rev. D} {\bf
  39} (Feb, 1989) 1020--1028}.

\bibitem{Mayle:1987as}
R.~Mayle, J.~R. Wilson, J.~R. Ellis, K.~A. Olive, D.~N. Schramm and
  G.~Steigman, \emph{{Constraints on Axions from SN 1987a}},
  \href{http://dx.doi.org/10.1016/0370-2693(88)91595-X}{\emph{Phys. Lett.} {\bf
  B203} (1988) 188--196}.

\bibitem{Mayle:1989yx}
R.~Mayle, J.~R. Wilson, J.~R. Ellis, K.~A. Olive, D.~N. Schramm and
  G.~Steigman, \emph{{Updated Constraints on Axions from SN 1987a}},
  \href{http://dx.doi.org/10.1016/0370-2693(89)91104-0}{\emph{Phys. Lett.} {\bf
  B219} (1989) 515}.

\bibitem{PhysRevLett.60.1793}
G.~Raffelt and D.~Seckel, \emph{Bounds on exotic-particle interactions from
  sn1987a}, \href{http://dx.doi.org/10.1103/PhysRevLett.60.1793}{\emph{Phys.
  Rev. Lett.} {\bf 60} (May, 1988) 1793--1796}.

\bibitem{Turner:1987by}
M.~S. Turner, \emph{{Axions from SN 1987a}},
  \href{http://dx.doi.org/10.1103/PhysRevLett.60.1797}{\emph{Phys. Rev. Lett.}
  {\bf 60} (1988) 1797}.

\bibitem{Tanabashi:2018oca}
{\scshape Particle Data Group} collaboration, M.~Tanabashi et~al.,
  \emph{{Review of Particle Physics}},
  \href{http://dx.doi.org/10.1103/PhysRevD.98.030001}{\emph{Phys. Rev. D} {\bf
  98} (2018) 030001}.

\bibitem{Bar:2019ifz}
N.~Bar, K.~Blum and G.~D'amico, \emph{{Is there a supernova bound on axions?}},
   \href{https://arxiv.org/abs/1907.05020}{{\tt 1907.05020}}.

\bibitem{Hook:2017psm}
A.~Hook and J.~Huang, \emph{{Probing axions with neutron star inspirals and
  other stellar processes}},
  \href{http://dx.doi.org/10.1007/JHEP06(2018)036}{\emph{JHEP} {\bf 06} (2018)
  036}, [\href{https://arxiv.org/abs/1708.08464}{{\tt 1708.08464}}].

\bibitem{Huang:2018pbu}
J.~Huang, M.~C. Johnson, L.~Sagunski, M.~Sakellariadou and J.~Zhang,
  \emph{{Prospects for axion searches with Advanced LIGO through binary
  mergers}}, \href{http://dx.doi.org/10.1103/PhysRevD.99.063013}{\emph{Phys.
  Rev.} {\bf D99} (2019) 063013}, [\href{https://arxiv.org/abs/1807.02133}{{\tt
  1807.02133}}].

\bibitem{Poddar:2019zoe}
T.~Kumar~Poddar, S.~Mohanty and S.~Jana, \emph{{Constraints on ultralight
  axions from compact binary systems}},
  \href{http://dx.doi.org/10.1103/PhysRevD.101.083007}{\emph{Phys. Rev. D} {\bf
  101} (2020) 083007}, [\href{https://arxiv.org/abs/1906.00666}{{\tt
  1906.00666}}].

\bibitem{Alford:2017rxf}
M.~G. Alford, L.~Bovard, M.~Hanauske, L.~Rezzolla and K.~Schwenzer,
  \emph{{Viscous Dissipation and Heat Conduction in Binary Neutron-Star
  Mergers}},
  \href{http://dx.doi.org/10.1103/PhysRevLett.120.041101}{\emph{Phys. Rev.
  Lett.} {\bf 120} (2018) 041101},
  [\href{https://arxiv.org/abs/1707.09475}{{\tt 1707.09475}}].

\bibitem{OPE}
O.~Benhar, \emph{From yukawa’s theory to the one-pion-exchange potential},
  2017.

\bibitem{Raffelt:2006cw}
G.~G. Raffelt, \emph{{Astrophysical axion bounds}},
  \href{http://dx.doi.org/10.1007/978-3-540-73518-2_3}{\emph{Lect. Notes Phys.}
  {\bf 741} (2008) 51--71}, [\href{https://arxiv.org/abs/hep-ph/0611350}{{\tt
  hep-ph/0611350}}].

\bibitem{Liu:2001iz}
B.~Liu, V.~Greco, V.~Baran, M.~Colonna and M.~Di~Toro, \emph{{Asymmetric
  nuclear matter: The Role of the isovector scalar channel}},
  \href{http://dx.doi.org/10.1103/PhysRevC.65.045201}{\emph{Phys. Rev.} {\bf
  C65} (2002) 045201}, [\href{https://arxiv.org/abs/nucl-th/0112034}{{\tt
  nucl-th/0112034}}].

\bibitem{Essick:2020flb}
R.~Essick, I.~Tews, P.~Landry, S.~Reddy and D.~E. Holz, \emph{{Direct
  Astrophysical Tests of Chiral Effective Field Theory at Supranuclear
  Densities}},  \href{https://arxiv.org/abs/2004.07744}{{\tt 2004.07744}}.

\bibitem{Fu:2008zzg}
W.-j. Fu, G.-h. Wang and Y.-x. Liu, \emph{{Electron Capture and Its Reverse
  Process in Hot and Dense Astronuclear Matter}},
  \href{http://dx.doi.org/10.1086/528361}{\emph{Astrophys. J.} {\bf 678} (2008)
  1517--1529}.

\bibitem{Leinson:2002bw}
L.~B. Leinson, \emph{{Direct Urca processes on nucleons in cooling neutron
  stars}}, \href{http://dx.doi.org/10.1016/S0375-9474(02)00991-0}{\emph{Nucl.
  Phys.} {\bf A707} (2002) 543--560},
  [\href{https://arxiv.org/abs/hep-ph/0207116}{{\tt hep-ph/0207116}}].

\bibitem{Roberts:2016mwj}
L.~F. Roberts and S.~Reddy, \emph{{Charged current neutrino interactions in hot
  and dense matter}},
  \href{http://dx.doi.org/10.1103/PhysRevC.95.045807}{\emph{Phys. Rev.} {\bf
  C95} (2017) 045807}, [\href{https://arxiv.org/abs/1612.02764}{{\tt
  1612.02764}}].

\bibitem{Alford:2019kdw}
M.~Alford, A.~Harutyunyan and A.~Sedrakian, \emph{{Bulk viscosity of baryonic
  matter with trapped neutrinos}},
  \href{http://dx.doi.org/10.1103/PhysRevD.100.103021}{\emph{Phys. Rev.} {\bf
  D100} (2019) 103021}, [\href{https://arxiv.org/abs/1907.04192}{{\tt
  1907.04192}}].

\bibitem{Perego:2019adq}
A.~Perego, S.~Bernuzzi and D.~Radice, \emph{{Thermodynamics conditions of
  matter in neutron star mergers}},
  \href{http://dx.doi.org/10.1140/epja/i2019-12810-7}{\emph{Eur. Phys. J.} {\bf
  A55} (2019) 124}, [\href{https://arxiv.org/abs/1903.07898}{{\tt
  1903.07898}}].

\bibitem{Most:2019onn}
E.~R. Most, L.~J. Papenfort, V.~Dexheimer, M.~Hanauske, H.~Stöcker and
  L.~Rezzolla, \emph{{On the Deconfinement Phase Transition in Neutron-Star
  Mergers}},
  \href{http://dx.doi.org/10.1140/epja/s10050-020-00073-4}{\emph{Eur. Phys. J.}
  {\bf A56} (2020) 59}, [\href{https://arxiv.org/abs/1910.13893}{{\tt
  1910.13893}}].

\bibitem{Horowitz:2005zv}
C.~J. Horowitz and A.~Schwenk, \emph{{The Virial equation of state of
  low-density neutron matter}},
  \href{http://dx.doi.org/10.1016/j.physletb.2006.05.055}{\emph{Phys. Lett.}
  {\bf B638} (2006) 153--159},
  [\href{https://arxiv.org/abs/nucl-th/0507064}{{\tt nucl-th/0507064}}].

\bibitem{Fore:2019wib}
B.~Fore and S.~Reddy, \emph{{Pions in hot dense matter and their astrophysical
  implications}},
  \href{http://dx.doi.org/10.1103/PhysRevC.101.035809}{\emph{Phys. Rev. C} {\bf
  101} (2020) 035809}, [\href{https://arxiv.org/abs/1911.02632}{{\tt
  1911.02632}}].

\bibitem{Ishizuka:1989ts}
N.~Ishizuka and M.~Yoshimura, \emph{{Axion and Dilaton Emissivity From Nascent
  Neutron Stars}}, \href{http://dx.doi.org/10.1143/PTP.84.233}{\emph{Prog.
  Theor. Phys.} {\bf 84} (1990) 233--250}.

\bibitem{glendenning2000compact}
N.~Glendenning, \emph{Compact Stars: Nuclear Physics, Particle Physics, and
  General Relativity}.
\newblock Astronomy and Astrophysics Library. Springer New York, 2000.

\bibitem{Paul:2018msp}
A.~Paul, D.~Majumdar and K.~Prasad~Modak, \emph{{Neutron star cooling via axion
  emission by nucleon–nucleon axion bremsstrahlung}},
  \href{http://dx.doi.org/10.1007/s12043-018-1702-2}{\emph{Pramana} {\bf 92}
  (2019) 44}, [\href{https://arxiv.org/abs/1801.07928}{{\tt 1801.07928}}].

\bibitem{Lee:2018lcj}
J.~S. Lee, \emph{{Revisiting Supernova 1987A Limits on Axion-Like-Particles}},
  \href{https://arxiv.org/abs/1808.10136}{{\tt 1808.10136}}.

\bibitem{Alford:2018lhf}
M.~G. Alford and S.~P. Harris, \emph{{Beta equilibrium in neutron star
  mergers}}, \href{http://dx.doi.org/10.1103/PhysRevC.98.065806}{\emph{Phys.
  Rev.} {\bf C98} (2018) 065806}, [\href{https://arxiv.org/abs/1803.00662}{{\tt
  1803.00662}}].

\bibitem{Iwamoto:1992jp}
N.~Iwamoto, \emph{{Nucleon-nucleon bremsstrahlung of axions and pseudoscalar
  particles from neutron star matter}},
  \href{http://dx.doi.org/10.1103/PhysRevD.64.043002}{\emph{Phys. Rev.} {\bf
  D64} (2001) 043002}.

\bibitem{Shapiro:1983du}
S.~L. Shapiro and S.~A. Teukolsky, \emph{{Black holes, white dwarfs, and
  neutron stars: The physics of compact objects}}.
\newblock 1983.

\bibitem{Fischer:2016cyd}
T.~Fischer, S.~Chakraborty, M.~Giannotti, A.~Mirizzi, A.~Payez and A.~Ringwald,
  \emph{{Probing axions with the neutrino signal from the next galactic
  supernova}}, \href{http://dx.doi.org/10.1103/PhysRevD.94.085012}{\emph{Phys.
  Rev.} {\bf D94} (2016) 085012}, [\href{https://arxiv.org/abs/1605.08780}{{\tt
  1605.08780}}].

\bibitem{Giannotti:2017hny}
M.~Giannotti, I.~G. Irastorza, J.~Redondo, A.~Ringwald and K.~Saikawa,
  \emph{{Stellar Recipes for Axion Hunters}},
  \href{http://dx.doi.org/10.1088/1475-7516/2017/10/010}{\emph{JCAP} {\bf 1710}
  (2017) 010}, [\href{https://arxiv.org/abs/1708.02111}{{\tt 1708.02111}}].

\bibitem{Berenji:2016jji}
B.~Berenji, J.~Gaskins and M.~Meyer, \emph{{Constraints on Axions and Axionlike
  Particles from Fermi Large Area Telescope Observations of Neutron Stars}},
  \href{http://dx.doi.org/10.1103/PhysRevD.93.045019}{\emph{Phys. Rev.} {\bf
  D93} (2016) 045019}, [\href{https://arxiv.org/abs/1602.00091}{{\tt
  1602.00091}}].

\bibitem{Sedrakian:2018kdm}
A.~Sedrakian, \emph{{Axion cooling of neutron stars. II. Beyond hadronic
  axions}}, \href{http://dx.doi.org/10.1103/PhysRevD.99.043011}{\emph{Phys.
  Rev. D} {\bf 99} (2019) 043011},
  [\href{https://arxiv.org/abs/1810.00190}{{\tt 1810.00190}}].

\bibitem{Sedrakian:2015krq}
A.~Sedrakian, \emph{{Axion cooling of neutron stars}},
  \href{http://dx.doi.org/10.1103/PhysRevD.93.065044}{\emph{Phys. Rev.} {\bf
  D93} (2016) 065044}, [\href{https://arxiv.org/abs/1512.07828}{{\tt
  1512.07828}}].

\bibitem{Beznogov:2018fda}
M.~V. Beznogov, E.~Rrapaj, D.~Page and S.~Reddy, \emph{{Constraints on
  Axion-like Particles and Nucleon Pairing in Dense Matter from the Hot Neutron
  Star in HESS J1731-347}},
  \href{http://dx.doi.org/10.1103/PhysRevC.98.035802}{\emph{Phys. Rev.} {\bf
  C98} (2018) 035802}, [\href{https://arxiv.org/abs/1806.07991}{{\tt
  1806.07991}}].

\bibitem{Lloyd:2019rxg}
S.~J. Lloyd, P.~M. Chadwick and A.~M. Brown, \emph{{Constraining the axion mass
  through gamma-ray observations of pulsars}},
  \href{http://dx.doi.org/10.1103/PhysRevD.100.063005}{\emph{Phys. Rev.} {\bf
  D100} (2019) 063005}, [\href{https://arxiv.org/abs/1908.03413}{{\tt
  1908.03413}}].

\bibitem{Yakovlev:2004iq}
D.~G. Yakovlev and C.~J. Pethick, \emph{{Neutron star cooling}},
  \href{http://dx.doi.org/10.1146/annurev.astro.42.053102.134013}{\emph{Ann.
  Rev. Astron. Astrophys.} {\bf 42} (2004) 169--210},
  [\href{https://arxiv.org/abs/astro-ph/0402143}{{\tt astro-ph/0402143}}].

\bibitem{Potekhin:2015qsa}
A.~Y. Potekhin, J.~A. Pons and D.~Page, \emph{{Neutron stars - cooling and
  transport}}, \href{http://dx.doi.org/10.1007/s11214-015-0180-9}{\emph{Space
  Sci. Rev.} {\bf 191} (2015) 239--291},
  [\href{https://arxiv.org/abs/1507.06186}{{\tt 1507.06186}}].

\bibitem{Haensel:1987zz}
P.~Haensel and A.~J. Jerzak, \emph{{Mean free paths of non-degenerate neutrinos
  in neutron star matter}}, {\emph{Astron. Astrophys.} {\bf 179} (1987)
  127--133}.

\bibitem{1979ApJ...230..859S}
R.~F. {Sawyer} and A.~{Soni}, \emph{{Transport of neutrinos in hot neutron-star
  matter}}, \href{http://dx.doi.org/10.1086/157146}{\emph{Astrophys. J.} {\bf
  230} (June, 1979) 859--869}.

\bibitem{Sawyer:1975js}
R.~F. Sawyer, \emph{{Neutrino Opacity of Neutron Star Matter}},
  \href{http://dx.doi.org/10.1103/PhysRevD.11.2740}{\emph{Phys. Rev.} {\bf D11}
  (1975) 2740}.

\bibitem{1982ApJ...253..816G}
B.~T. {Goodwin} and C.~J. {Pethick}, \emph{{Transport properties of degenerate
  neutrinos in dense matter}},
  \href{http://dx.doi.org/10.1086/159684}{\emph{Astrophys. J.} {\bf 253} (Feb.,
  1982) 816--838}.

\bibitem{1994ARep...38..247L}
K.~P. {Levenfish} and D.~G. {Yakovlev}, \emph{{Specific heat of neutron star
  cores with superfluid nucleons}}, {\emph{Astronomy Reports} {\bf 38} (Mar,
  1994) 247--251}.

\bibitem{Maslov:2015wba}
K.~A. Maslov, E.~E. Kolomeitsev and D.~N. Voskresensky, \emph{{Relativistic
  Mean-Field Models with Scaled Hadron Masses and Couplings: Hyperons and
  Maximum Neutron Star Mass}},
  \href{http://dx.doi.org/10.1016/j.nuclphysa.2016.03.011}{\emph{Nucl. Phys.}
  {\bf A950} (2016) 64--109}, [\href{https://arxiv.org/abs/1509.02538}{{\tt
  1509.02538}}].

\bibitem{Li:2018lpy}
B.-A. Li, B.-J. Cai, L.-W. Chen and J.~Xu, \emph{{Nucleon Effective Masses in
  Neutron-Rich Matter}},
  \href{http://dx.doi.org/10.1016/j.ppnp.2018.01.001}{\emph{Prog. Part. Nucl.
  Phys.} {\bf 99} (2018) 29--119},
  [\href{https://arxiv.org/abs/1801.01213}{{\tt 1801.01213}}].

\bibitem{PhysRevLett.53.1198}
N.~Iwamoto, \emph{Axion emission from neutron stars},
  \href{http://dx.doi.org/10.1103/PhysRevLett.53.1198}{\emph{Phys. Rev. Lett.}
  {\bf 53} (Sep, 1984) 1198--1201}.

\bibitem{Stoica:2009zh}
S.~Stoica, B.~Pastrav, J.~E. Horvath and M.~P. Allen, \emph{{Pion mass effects
  on axion emission from neutron stars through NN bremsstrahlung processes}},
  \href{http://dx.doi.org/10.1016/j.nuclphysa.2009.07.007,
  10.1016/j.nuclphysa.2009.12.001}{\emph{Nucl. Phys.} {\bf A828} (2009)
  439--449}, [\href{https://arxiv.org/abs/0906.3134}{{\tt 0906.3134}}].

\bibitem{Hanauske:2016gia}
M.~Hanauske, K.~Takami, L.~Bovard, L.~Rezzolla, J.~A. Font, F.~Galeazzi et~al.,
  \emph{{Rotational properties of hypermassive neutron stars from binary
  mergers}}, \href{http://dx.doi.org/10.1103/PhysRevD.96.043004}{\emph{Phys.
  Rev.} {\bf D96} (2017) 043004}, [\href{https://arxiv.org/abs/1611.07152}{{\tt
  1611.07152}}].

\bibitem{Hanauske:2019czl}
M.~Hanauske, L.~Bovard, E.~Most, J.~Papenfort, J.~Steinheimer, A.~Motornenko
  et~al., \emph{{Detecting the Hadron-Quark Phase Transition with Gravitational
  Waves}}, \href{http://dx.doi.org/10.3390/universe5060156}{\emph{Universe}
  {\bf 5} (2019) 156}.

\bibitem{Hanauske:2019vsz}
M.~Hanauske, L.~Bovard, J.~Steinheimer, A.~Motornenko, V.~Vovchenko, S.~Schramm
  et~al., \emph{{MAGIC - how MAtter’s extreme phases can be revealed in
  Gravitational wave observations and in relativistic heavy Ion Collision
  experiments}},
  \href{http://dx.doi.org/10.1088/1742-6596/1271/1/012023}{\emph{J. Phys. Conf.
  Ser.} {\bf 1271} (2019) 012023}.

\bibitem{Alford:2019qtm}
M.~G. Alford and S.~P. Harris, \emph{{Damping of density oscillations in
  neutrino-transparent nuclear matter}},
  \href{http://dx.doi.org/10.1103/PhysRevC.100.035803}{\emph{Phys. Rev.} {\bf
  C100} (2019) 035803}, [\href{https://arxiv.org/abs/1907.03795}{{\tt
  1907.03795}}].

\bibitem{Keil:1996ju}
W.~Keil, H.-T. Janka, D.~N. Schramm, G.~Sigl, M.~S. Turner and J.~R. Ellis,
  \emph{{A Fresh look at axions and SN-1987A}},
  \href{http://dx.doi.org/10.1103/PhysRevD.56.2419}{\emph{Phys. Rev.} {\bf D56}
  (1997) 2419--2432}, [\href{https://arxiv.org/abs/astro-ph/9612222}{{\tt
  astro-ph/9612222}}].

\bibitem{PhysRevD.52.1780}
G.~Raffelt and D.~Seckel, \emph{Self-consistent approach to neutral-current
  processes in supernova cores},
  \href{http://dx.doi.org/10.1103/PhysRevD.52.1780}{\emph{Phys. Rev. D} {\bf
  52} (Aug, 1995) 1780--1799}.

\bibitem{Hanhart:2000ae}
C.~Hanhart, D.~R. Phillips and S.~Reddy, \emph{{Neutrino and axion emissivities
  of neutron stars from nucleon-nucleon scattering data}},
  \href{http://dx.doi.org/10.1016/S0370-2693(00)01382-4}{\emph{Phys. Lett.}
  {\bf B499} (2001) 9--15}, [\href{https://arxiv.org/abs/astro-ph/0003445}{{\tt
  astro-ph/0003445}}].

\bibitem{Bartl:2016iok}
A.~Bartl, R.~Bollig, H.-T. Janka and A.~Schwenk, \emph{{Impact of
  Nucleon-Nucleon Bremsstrahlung Rates Beyond One-Pion Exchange}},
  \href{http://dx.doi.org/10.1103/PhysRevD.94.083009}{\emph{Phys. Rev.} {\bf
  D94} (2016) 083009}, [\href{https://arxiv.org/abs/1608.05037}{{\tt
  1608.05037}}].

\bibitem{Balkin:2020dsr}
R.~Balkin, J.~Serra, K.~Springmann and A.~Weiler, \emph{{The QCD Axion at
  Finite Density}},  \href{https://arxiv.org/abs/2003.04903}{{\tt 2003.04903}}.

\bibitem{Horowitz:2003yx}
C.~J. Horowitz and M.~A. Perez-Garcia, \emph{{Realistic neutrino opacities for
  supernova simulations with correlations and weak magnetism}},
  \href{http://dx.doi.org/10.1103/PhysRevC.68.025803}{\emph{Phys. Rev.} {\bf
  C68} (2003) 025803}, [\href{https://arxiv.org/abs/astro-ph/0305138}{{\tt
  astro-ph/0305138}}].

\bibitem{Fischer:2018kdt}
T.~Fischer, G.~Guo, A.~A. Dzhioev, G.~Martínez-Pinedo, M.-R. Wu, A.~Lohs
  et~al., \emph{{Neutrino signal from proto-neutron star evolution: Effects of
  opacities from charged-current-neutrino interactions and inverse neutron
  decay}}, \href{http://dx.doi.org/10.1103/PhysRevC.101.025804}{\emph{Phys.
  Rev.} {\bf C101} (2020) 025804},
  [\href{https://arxiv.org/abs/1804.10890}{{\tt 1804.10890}}].

\bibitem{Ardevol-Pulpillo:2018btx}
R.~Ardevol-Pulpillo, H.~T. Janka, O.~Just and A.~Bauswein, \emph{{Improved
  Leakage-Equilibration-Absorption Scheme (ILEAS) for Neutrino Physics in
  Compact Object Mergers}},
  \href{http://dx.doi.org/10.1093/mnras/stz613}{\emph{Mon. Not. Roy. Astron.
  Soc.} {\bf 485} (2019) 4754--4789},
  [\href{https://arxiv.org/abs/1808.00006}{{\tt 1808.00006}}].

\bibitem{Perego:2014qda}
A.~Perego, E.~Gafton, R.~Cabezón, S.~Rosswog and M.~Liebendörfer,
  \emph{{MODA: a new algorithm to compute optical depths in multidimensional
  hydrodynamic simulations}},
  \href{http://dx.doi.org/10.1051/0004-6361/201423755}{\emph{Astron.
  Astrophys.} {\bf 568} (2014) A11},
  [\href{https://arxiv.org/abs/1403.1297}{{\tt 1403.1297}}].

\bibitem{Galeazzi:2013mia}
F.~Galeazzi, W.~Kastaun, L.~Rezzolla and J.~A. Font, \emph{{Implementation of a
  simplified approach to radiative transfer in general relativity}},
  \href{http://dx.doi.org/10.1103/PhysRevD.88.064009}{\emph{Phys. Rev.} {\bf
  D88} (2013) 064009}, [\href{https://arxiv.org/abs/1306.4953}{{\tt
  1306.4953}}].

\bibitem{Sekiguchi:2012uc}
Y.~Sekiguchi, K.~Kiuchi, K.~Kyutoku and M.~Shibata, \emph{{Current Status of
  Numerical-Relativity Simulations in Kyoto}},
  \href{http://dx.doi.org/10.1093/ptep/pts011}{\emph{PTEP} {\bf 2012} (2012)
  01A304}, [\href{https://arxiv.org/abs/1206.5927}{{\tt 1206.5927}}].

\bibitem{Rosswog:2003rv}
S.~Rosswog and M.~Liebendoerfer, \emph{{High resolution calculations of merging
  neutron stars. 2: Neutrino emission}},
  \href{http://dx.doi.org/10.1046/j.1365-8711.2003.06579.x}{\emph{Mon. Not.
  Roy. Astron. Soc.} {\bf 342} (2003) 673},
  [\href{https://arxiv.org/abs/astro-ph/0302301}{{\tt astro-ph/0302301}}].

\bibitem{1999JCoAM.109..281M}
A.~{Mezzacappa} and O.~E.~B. {Messer}, \emph{{Neutrino transport in core
  collapse supernovae.}}, {\emph{Journal of Computational and Applied
  Mathematics} {\bf 109} (Sept., 1999) 281--319}.

\bibitem{Kaminker:2016ayg}
A.~D. Kaminker, D.~G. Yakovlev and P.~Haensel, \emph{{Theory of neutrino
  emission from nucleon–hyperon matter in neutron stars: angular integrals}},
  \href{http://dx.doi.org/10.1007/s10509-016-2854-5}{\emph{Astrophys. Space
  Sci.} {\bf 361} (2016) 267}, [\href{https://arxiv.org/abs/1607.05265}{{\tt
  1607.05265}}].

\bibitem{Giannotti:2005tn}
M.~Giannotti and F.~Nesti, \emph{{Nucleon-nucleon Bremsstrahlung emission of
  massive axions}},
  \href{http://dx.doi.org/10.1103/PhysRevD.72.063005}{\emph{Phys. Rev.} {\bf
  D72} (2005) 063005}, [\href{https://arxiv.org/abs/hep-ph/0505090}{{\tt
  hep-ph/0505090}}].

\bibitem{Dent:2012mx}
J.~B. Dent, F.~Ferrer and L.~M. Krauss, \emph{{Constraints on Light Hidden
  Sector Gauge Bosons from Supernova Cooling}},
  \href{https://arxiv.org/abs/1201.2683}{{\tt 1201.2683}}.

\bibitem{Carenza:2019pxu}
P.~Carenza, T.~Fischer, M.~Giannotti, G.~Guo, G.~Martínez-Pinedo and
  A.~Mirizzi, \emph{{Improved axion emissivity from a supernova via
  nucleon-nucleon bremsstrahlung}},
  \href{http://dx.doi.org/10.1088/1475-7516/2019/10/016}{\emph{JCAP} {\bf 1910}
  (2019) 016}, [\href{https://arxiv.org/abs/1906.11844}{{\tt 1906.11844}}].

\end{thebibliography}\endgroup

\end{document}